\documentclass[dvipsnames]{aa}
\usepackage{graphicx}
\usepackage[varg]{txfonts}
\usepackage{scalerel}
\usepackage{hyperref}
\hypersetup{colorlinks=true, citecolor=PineGreen}
\usepackage[capitalize]{cleveref}
\usepackage{xcolor}
\usepackage{xspace}
\usepackage[detect-weight=true, detect-family=true]{siunitx}
\usepackage{booktabs}
\usepackage{multicol}
\usepackage[normalem]{ulem}

\DeclareSIUnit\magnitude{mag}
\DeclareSIUnit\coreh{core\textrm{-}h}
\DeclareSIUnit\msol{\textrm{M}_{\odot}}
\DeclareSIUnit\mach{Ma}
\DeclareSIUnit\erg{erg}

\newcommand{\bvf}{BVF\@\xspace}
\newcommand{\vconv}{v_{\mathrm{conv}}}
\newcommand{\edel}{EM19\xspace}
\newcommand{\figr}[1]{figure~#1}
\newcommand{\figrs}[1]{figures~#1}
\newcommand{\eqr}[1]{equation~#1}
\newcommand{\dd}{\mathrm{d}}
\newcommand{\DD}{\mathrm{D}}
\newcommand{\thse}{\vartheta_{\mathrm{hse}}}
\newcommand{\ma}{\mathrm{Ma}}
\newcommand{\fma}{f_{\ma}}
\newcommand{\gma}{\ma_{gy}}
\newcommand{\ausmpup}{AUSM$^{+}$-up\xspace}
\newcommand{\roe}{Roe solver\xspace}
\newcommand{\obvf}{\frac{2\pi}{N_0}}
\newcommand{\slh}{SLH\@\xspace}

\newcommand{\cflx}[1]{\ensuremath{\mathrm{CFL}_{\mathrm{#1}}}}
\newcommand{\dtx}[1]{\delta t_{\scaleto{\mathrm{#1}}{5pt}}}
\newcommand{\tcross}{t_{\mathrm{cross}}}
\newcommand{\tconv}{\overline{\tau}_{\mathrm{conv}}}
\newcommand{\tdiff}{\tau_{\mathrm{diff}}}

\newcommand{\ii}{i}
\newcommand{\Aa}{\mathcal{A}}
\newcommand{\omo}{\omega|_{\vec k_0}}
\newcommand{\omot}{\omo t}
\newcommand{\Dq}[1]{\Delta #1}

\begin{document}
\title{Fully compressible simulations of waves and core convection\\ in main-sequence stars}
\author{L. Horst\inst{1}\and
        P. V. F. Edelmann\inst{2,3}\and
        R. Andr\'assy\inst{1}\and
        F. K. R\"{o}pke\inst{1,4}\and
        D. M. Bowman\inst{5}\and
        C. Aerts\inst{5,6,7}\and
        R. P. Ratnasingam\inst{2}}
\titlerunning{Fully compressible simulations of waves and core convection in main-sequence stars}
\authorrunning{Horst et al.}

\institute{%
  Heidelberger Institut f\"{u}r Theoretische Studien,
  Schloss-Wolfsbrunnenweg 35, 69118 Heidelberg, Germany\\
  \email{leonhard.horst@h-its.org}
  \and
  School of Mathematics, Statistics and Physics,
  Newcastle University,
  Newcastle upon Tyne,
  NE1 7RU,
  UK
  \and
  X Computational Physics (XCP) Division and Center for Theoretical Astrophysics (CTA),
  Los Alamos National Laboratory,
  Los Alamos,
  NM 87545,
  USA
  \and
  Zentrum f\"ur Astronomie der Universit\"at Heidelberg,
  Institut f\"ur Theoretische Astrophysik,
  Philosophenweg 12,
  69120 Heidelberg,
  Germany
  \and
  Instituut voor Sterrenkunde,
  KU Leuven,
  Celestijnenlaan 200D,
  3001, Leuven,
  Belgium
  \and
  Department of Astrophysics,
  IMAPP,
  Radboud University Nijmegen,
  PO Box 9010,
  6500 GL, Nijmegen,
  The Netherlands
  \and
  Max Planck Institute for Astronomy,
  K\"onigstuhl 17,
  69117 Heidelberg,
  Germany
}

\date{Received 20 January 2020 / accepted 3 July 2020} 
\abstract{%
  \textbf{Context.} Recent, nonlinear simulations of wave generation and
  propagation in full-star models have been carried out in the anelastic
  approximation using spectral methods. Although it makes long time steps
  possible, this approach excludes the physics of sound waves completely
  and requires rather high artificial viscosity and thermal diffusivity
  for numerical stability. A direct comparison with observations is thus
  limited.
  \\
  \textbf{Aims.} We explore the capabilities of our compressible
  multidimensional Seven-League Hydro (\slh) code to simulate stellar oscillations. 
  \\
  \textbf{Methods.} We compare some fundamental properties of internal gravity
  and pressure waves in 2D SLH simulations to linear wave theory using two test
  cases: (1) an interval gravity wave packet in the Boussinesq limit and (2) a
  realistic \SI{3}{\msol} stellar model with a convective core and a radiative
  envelope. Oscillation properties of the stellar model are also discussed in
  the context of observations.
  \\
  \textbf{Results.} Our tests show that specialized low-Mach techniques are
  necessary when simulating oscillations in stellar interiors. Basic properties
  of internal gravity and pressure waves in our simulations are in good
  agreement with linear wave theory. As compared to anelastic simulations of
  the same stellar model, we can follow internal gravity waves of much lower
  frequencies. The temporal frequency spectra of velocity and temperature are
  flat and compatible with the observed spectra of massive stars.
  \\
  \textbf{Conclusion.}
  The low-Mach compressible approach to hydrodynamical simulations of stellar
  oscillations is promising. Our simulations are less dissipative and require
  less luminosity boosting than comparable spectral simulations. The
  fully-compressible approach allows for the coupling of gravity and
  pressure waves in the outer convective envelopes of evolved stars to be
  studied in the future.
  }

\maketitle

\section{Introduction}\label{sec:introduction}

The study of the excitation and propagation of waves within stars has greatly
helped to shape stellar structure and evolution theory over the last century.
Today, this area of astronomy is called asteroseismology, and it includes the
study of oscillations across the Hertzsprung--Russell diagram. For example,
pressure modes (p-modes) provide important constraints on the envelopes of
stars. Modes of a consecutive radial order ($n$) and the same angular degree
($\ell$) have a characteristic frequency separation known as the large
frequency separation, which is sensitive to the average density of a star
\citep{aerts2010a}.  This application of asteroseismology using p-modes has
been extremely successful for low- and intermediate-mass stars
\citep{chaplin2013c, hekker2017a, garcia2019a}.  Specifically, the measurement
of envelope rotation using rotationally-split pressure modes has facilitated
the discovery that stars with masses of about \SI{2}{\msol} have approximately
rigid interior rotation profiles \citep[see][]{kurtz2014, saio2015b,
vanreeth2016a, vanreeth2018a}. Hence, current angular momentum theory already needs
significant improvement on the main sequence (MS) \citep{aerts2019b}. 

For later evolutionary stages, including subgiant, red giant, and red clump
stars, pulsations that behave as gravity modes (g-modes) near the core and
as p-modes near the surface have been detected in thousands of stars
\citep[see][]{beck2011a}. These ``mixed modes'' can be used to distinguish
different stages of nuclear burning \citep{bedding2011}. Hence, understanding
pressure modes is not only crucial for measuring interior properties of main
sequence stars, but also for post-MS stars \citep[see, e.g.,][]{beck2012b,
mosser2012c}.

On the upper main sequence, stars with spectral types O and B
($M>\SI{3}{\msol}$) observed in \SI{}{\mu\magnitude} precision space photometry
show coherent opacity-driven p- or g-modes as well as stochastic variability
caused by internal gravity waves. This occurs in slowly pulsating B (SPB) stars
with masses between \SI{3}{\msol} and \SI{9}{\msol} \citep{papics2017a,
bowman2019a, pedersen2020a}, in $\beta\,$Cep stars with masses between
\SI{8}{\msol} and \SI{25}{\msol} \citep{briquet2011a, burssens2019a}, and in
young and evolved O-type dwarfs and blue supergiants with masses up to $\sim\!
\SI{50}{\msol}$ \citep{buysschaert2015a, bowman2019b, pedersen2019a, bowman2020}. This
overwhelming observational evidence from CoRoT, {\it Kepler}/K2, and TESS
photometry motivated the development and study of our simulation setup.

The typical numerical approach to model the excitation of waves generated
within stars is to solve the Navier--Stokes equations in the anelastic
approximation \citep[e.g.,][]{rogers2013a,alvan2014a,edelmann2019a}. This method
allows for large time steps while still being a mostly explicit method and is
therefore computationally efficient. While being suitable to simulate internal
gravity waves (IGWs) where gravity is the dominant restoring force, some
important phenomena, such as the excitation of p-modes, cannot be followed.
Furthermore, common numerical methods to solve the equations in the anelastic
approximation require to introduce an artificial viscosity to achieve numerical
stability. To balance the effect of high viscosity, the stellar luminosity and
the thermal diffusivity have to be increased by orders of magnitude. This leads
to a damping of the waves, especially in the low frequency regime.

Some of these drawbacks are avoided or reduced by performing compressible
simulations of stellar interiors. Here, the full Navier--Stokes equations are
solved, most commonly in the finite volume approach. This includes the physics
of sound waves and allows the luminosity and the thermal diffusivity to be kept
much closer to stellar values. Viscosity is implicitly introduced by the
numerical scheme, and is lower than the viscosity typically used by spectral
codes (nevertheless still orders of magnitudes higher than the astrophysical
value). This is why we commonly speak of solving the Euler equations, which follow from the
Navier--Stokes equations without an explicit viscosity term, in this context.
On the other hand, this kind of simulations comes with higher computational
costs compared to their anelastic counterparts.  That compressible simulations
show excited IGWs and p-modes has already been reported in the past. For
example, \citet{meakin2006a, meakin2007b} show a spectrum of the velocity for a
simulation of carbon and oxygen burning in a 3D wedge geometry. They compare
the form of a wave with predominately g-mode character and a coupled p- and
g-mode to the predictions from linear theory and find good agreement for both.
Also \citet{herwig2006a} indicate the excitation of p- and g-modes for the case
of He-shell burning. They find the frequency of the different modes to be
independent of resolution and boosting.

These prior studies focus on the effects of convective boundary mixing rather
than the physics of waves. Thus, the computational domains only contain small
parts of the radiative zone below and above the convective shells in evolved
stars. The frequency spectrum might therefore differ considerably compared to a
full star model and comparison to observations is difficult. Also, the analysis
of internal waves mainly consists of computing the resulting spectra without
further investigations. 

Therefore, to further assess the advantages of compressible simulations, we use
our finite volume Seven-League Hydro (\slh) code (for a description
see Sect.~\ref{sec:slh}) to examine in more detail the properties of excited
waves. To ease the validation and comparison of our results, we repeat the
simulation of a \SI{3}{\msol} zero-age main-sequence (ZAMS) model by
\citet{edelmann2019a} (\edel hereafter). For their simulation, \edel applied
the anelastic approximation and a comparison between these two approaches is
therefore possible. We note that many of the diagnostics presented in the main
part of this work originate from \edel.

Because 3D simulations of an entire stellar model are costly even on today's
supercomputer facilities, we use 2D simulations in this initial verification
experiment. The 2D approach allows us to cover almost the entire stellar radius
in our simulation domain, excluding only a small part in the core and the
outermost layers, and to follow the evolution for an extended period of time.
The computational costs are low enough to run the simulations on the local
computer cluster of the Heidelberg Institute for Theoretical Studies
(HITS), Germany. Convection is known to be only accurately described in three
spatial dimensions (see, e.g., \citealp{meakin2007b} for a comparison of 2D and
3D oxygen burning). However, as pointed out by \edel, the results of their 3D
simulation are compatible with those of a 2D simulation by \citet{rogers2013a}
for a different, but similar \SI{3}{\msol} model.  This indicates that 2D
simulations may still serve useful results for excited wave properties despite
their reduced dimensionality.

This work is intended as a proof-of-concept: We present and validate in detail
the results of simulating IGWs with the compressible hydrodynamics code \slh.
The lack of the third dimension may not allow for a full quantitative comparison
with observations, but some characteristics of 2D simulations are still
expected to match observations of the integrated variability at the surface in
a qualitative way.  The main questions we aim to answer are therefore: Can the
excited waves be identified as pressure waves and IGWs and do they comply with
the theoretical expectations?  Additionally, we compare the excited wave
spectra to previous 2D work by \citet{rogers2013a} and perform qualitative
comparisons to observations of frequency spectra of massive stars.

The paper is organized as follows: In \cref{sec:slh} we give a brief overview on
the numerical methods that are used in the \slh code. In \cref{sec:boussinesq}
we describe and apply a simple test setup to benchmark the capability of \slh to
treat IGWs\@. The 2D results for the \SI{3}{\msol} ZAMS model are discussed in
\cref{sec:results}. \cref{sec:conclusion} summarizes the most important aspects
and gives an outlook for future simulations.

\section{The SLH code}\label{sec:slh}

We use the \slh code for all simulations presented in this paper. It was
developed initially by \citet{miczek2013a}  and solves the fully compressible
Euler equations in a finite volume framework.  It allows us to choose between
an ideal gas and a more general equation of state that includes contributions
from radiation and electron degeneracy \citep{timmes2000a}. The hydro solver is
coupled to a nuclear reaction network \citep{edelmann2014a}. Radiation is
treated in the diffusion limit, which is appropriate in the optically thick
regions of the star covered here. The implemented mapping procedure between the
uniform Cartesian computational grid and a general curvilinear physical grid
introduces flexibility regarding grid geometry. Our implementation of the
mapping is based on \citet{kifonidis2012a} and \citet{colella2011a}, examples for
curvilinear grids can be found for example in \citet{calhoun2008a}.

The \slh code was developed with a focus on flows at very low Mach numbers. One
complication in this regime is the need of specialized flux scheme as common
approaches are dominated by artificial dissipation (see, e.g., 
\citealp{miczek2015a, barsukow2017a}). For the simulations presented here, we
use the \ausmpup solver \citep{liou2006a}. It splits the flux function into a
pressure and an advective part and has improved low-Mach capabilities. An
artificial diffusive component for both parts is introduced for stability. To
avoid divergence at very low Mach numbers, the scaling of the diffusion terms
is limited by a cutoff Mach number $\mathit{Mcut}$, which is a free parameter.
For technical reasons, a separate cutoff number $\mathit{Mcut}_\mathit{pdiff}$
for the pressure diffusion term is used in \slh.  In the simulations presented
here, we use $\mathit{Mcut}=\num{e-10}$ and $\mathit{Mcut}_\mathit{pdiff}=0.1$.
For lower values of $\mathit{Mcut}_\mathit{pdiff}$, the convergence rate of the
implicit solver decreases considerably.

Due to the large pressure gradient in stars, the \ausmpup solver quickly
destroys the hydrostatic stratification. To resolve this problem, \slh applies
a variant of the Cargo--Leroux well-balancing technique. The basic
idea is to remove the static background stratification from the conserved
variables before they enter the numerical flux function. This considerably
reduces the ``effective'' pressure gradient. The scheme was developed by
\citet{cargo1994a} for one-dimensional setups. \citet{edelmann2014a} describes
how this approach can be extended to multi-dimensional setups.

The flexible modular design of \slh facilitates the implementation of new
developments such that newly published flux functions or reconstruction schemes
can be easily implemented and tested. This makes \slh well suited to push
hydrodynamical simulations toward low Mach numbers. For the 2D simulation of
core hydrogen burning at the ZAMS presented in \cref{sec:results},
mixing-length theory predicts convection at Mach numbers around \num{e-4} which
calls for a numerical scheme optimized for slow flows. For a more detailed
description of the code we refer the reader to \citet{miczek2013a},
\citet{miczek2015a} and \citet{edelmann2014a}. 

At low Mach numbers, a common drawback of conventional \emph{explicit time
stepping} methods is that for stability the maximum possible time step size
$\dtx{CFL_{uc}}$ is limited to the acoustic Courant--Friedrich--Lewy (CFL)
criterion
\begin{align}
  \dtx{\cflx{uc}} = \frac{\cflx{uc}}{N_\mathrm{dim}}
        \min \left(\frac{\Delta x}{|u|+c_{\text{sound}}}\right).
  \label{eq:cfluc}
\end{align}
Here, $N_\mathrm{dim}$ is the number of dimensions, $\Delta x$ is the grid
spacing in the different coordinate directions, $u$ is the fluid velocity and
$c_{\text{sound}}$ is the speed of sound. \cflx{uc} is a dimensionless number
which needs to be smaller than unity. The factor $1/N_\mathrm{dim}$ in
combination with the minimum over all cells for the cell crossing time in all
directions gives a lower limit for the time step.  For low-Mach flows, $u\ll
c_{\text{sound}}$ and the time step size is dominated by the speed of sound.
As a consequence, many small time steps are needed to resolve the fluid flow.
Although the computational costs for one explicit time step are low, the large
number of necessary steps makes explicit time stepping inefficient in this
regime.

The great advantage of \emph{implicit time stepping} is that there is no
restriction for the step size required for stability. The main constraint
arises from the question of how well the flow is to be resolved in time. This
leads to the ``advective'' \,\cflx{u} criterion which results from
\cref{eq:cfluc} by omitting the speed of sound:
\begin{align}
  \dtx{\cflx{u}} = \frac{\cflx{u}}{N_\mathrm{dim}} \min \left(\frac{\Delta x}{|u|}\right).
  \label{eq:cflu}
\end{align}
For illustration purposes, setting $\cflx{u}=0.5$ corresponds to a time step which does
not allow the fluid to cross more than half of a cell per step.

Implicit time integration, however, requires to solve nonlinear systems of
equations in each step. This greatly increases the computational costs for one
implicit time step compared to explicit time stepping.

In \slh, the family of Explicit first stage, Singly Diagonally Implicit
Runge-Kutta (ESDIRK) time steppers is implemented following the description of
\citet{hosea1996a} and \citet{kennedy2001a}. For the simulations in this paper,
the ESDIRK23 scheme is applied. It  consists of three computing stages and is
up to second order accurate in time. This results in two nonlinear
systems of equations. \slh applies the Newton-Raphson method which finds their
solution in an iterative way. Each iteration itself needs the solution of a
linear system of equations. For these, \slh offers a variety of
different methods (e.g., Krylov-Subspace schemes and multigrid solvers).

At low Mach numbers, the increased computational costs per time step are
overcompensated by the benefit of a larger step size. Numerical tests with \slh
imply that implicit time stepping becomes more efficient than explicit time
stepping at Mach numbers smaller than about 0.1 to 0.01.

Apart from accuracy requirements, the length of implicit time steps is
also limited by the Newton-Raphson solver, which may converge slowly or
even diverge if the time step becomes too long. This limit depends on
the problem solved and on details of the numerical scheme and needs
to be determined experimentally.

The choice of the time step size for the 2D simulation presented in
\cref{sec:results} is discussed in \cref{subsec:timestepping} along with an
efficiency comparison between implicit and explicit time stepping.

\section{Testing internal gravity waves in SLH}\label{sec:boussinesq}

In this section, we scrutinize the capability of \slh to correctly reproduce
the propagation of IGWs. The base setup is a 2D Cartesian domain containing an
isothermal ideal gas in a hydrostatic stratification. A wave packet of small
amplitude is evolved for several oscillation periods. The group velocity and
the change of the wave shape are then extracted from the simulation and
compared to the theoretical prediction which follows from the linear Boussinesq
approximation. With this simple but well-defined test setup we verify whether
\slh is able to reproduce the prediction accurately enough before applying it
to the more complex case of a realistic stellar profile in \cref{sec:results}.

\subsection{Boussinesq IGW setup}\label{subsec:boussinesqsetup}

The theoretical basics of the benchmark test are presented in
\cref{appendix:busigw} and follow \citet{sutherland2010}.  The actual
experimental \slh setup closely follows the idea of \citet{miczek2013a} and is
extended up to Mach numbers of $\ma=\num{e-2}$. The basic parameters of the
setup are listed in \cref{tab:bigw}.
\begin{table}[htpb]
  \centering
  \caption{Parameters of the Boussinesq IGW simulation.}
  \label{tab:bigw}
  \begin{tabular}{ll}
    \toprule
    temperature            & $T_0 = \SI{300}{\kelvin}$                                          \\
    mean mol.\ weight      & $\mu = 1$                                                          \\
    adiabatic index        & $\gamma = 5/3$                                                     \\
    density at zero height & $\rho_0 = \SI{1}{\g\per\cm\cubed}$                                 \\
    gravity                & $\vec g = \SI{-e3}{\centi\metre\per\second\squared} \vec e_y$      \\
    \midrule
    resolution             & $288(x)\times 288(y)$                                              \\
    domain                 & $\left[0,\left(2\pi/|k_x|\right)\right]\times\left[y_1,y_2\right]$ \\
    boundary conditions    & x-direction: periodic                                              \\
                           & y-direction: constant ghost cells                                  \\
    time stepping          & ESDIRK23                                                           \\
    time step size         & $\delta t = \frac{1}{20} \obvf$                                    \\
    reconstruction         & linear, second order in space                                      \\
    \bottomrule
  \end{tabular}
\end{table}

Assuming the ideal gas law $p = \rho \mathcal{R}/\mu T$, where $\mathcal{R}$ is
the universal gas constant, the profiles for density, pressure, and potential
temperature in hydrostatic equilibrium are given by
\begin{align}
  \rho_{\mathrm{hse}} &= \rho_0 \exp\left( -\frac{y}{H_p} \right), \label{eq:irho}\\
  p_{\mathrm{hse}} &= \rho_0 \frac{\mathcal{R}}{\mu} T_0 \exp\left( -\frac{y}{H_p} \right),
    \label{req:ipres} \\
  \vartheta_{\mathrm{hse}} &= T_0 \exp\left( \frac{y}{H_p}\, \frac{\gamma-1}{\gamma}
    \right), \label{eq:ipte}
\end{align}
where the pressure scale height $H_p$ is defined as
\begin{align}
  H_p^{-1} = -\frac{\partial \ln p}{\partial y} = \frac{g\mu}{\mathcal{R} T_0}.
\end{align}
The Brunt--V\"ais\"al\"a frequency (\bvf) according to \cref{eq:bvf} is
spatially constant and reads
\begin{align}
  N_0 = \sqrt{\frac{g}{H_p}\, \frac{\gamma - 1}{\gamma}}.
  \label{eq:bvfideal}
\end{align}
We perturb this hydrostatic stratification with a monochromatic internal
gravity wave packet. The wavelength $\lambda = 2\pi/|\vec k|=\beta H_p$ of the
packet is set in terms of the fraction $\beta$ of the pressure scale height and
is inclined by \SI{-60}{\degree} with respect to the horizontal direction. We
therefore have
\begin{align}
  |\vec k| = \frac{2\pi}{\beta H_p}\, ,\quad\
  \theta = -\frac{\SI{60}{\degree}}{\SI{180}{\degree}}\pi
\end{align}
and
\begin{align}
  k_x = |\vec k|\cos\theta,\quad k_y = |\vec k|\sin\theta.
  \label{eq:kabs}
\end{align}

We set the vertical domain of the Cartesian box such that it corresponds to
\num{13} times the wavelength $\lambda$
\begin{align}
  y_1 = -5 \beta H_p,\quad y_2 = 8 \beta H_p
  \label{eq:ydomain}
\end{align}
in order to provide enough space for the wave to move upward in y-direction.
The horizontal extent is set to contain one horizontal wavelength $\lambda_x =
2\pi/|k_x|$, which, in combination with periodic boundary conditions, allows
for the plane wave approach. At the top and bottom boundary we apply a layer of
two cells which are filled with the hydrostatic initial condition but kept
constant in time (constant ghost cells). The amplitude for the vertical
velocity component $A_v$ is modulated by a Gaussian function according to
\begin{align}
  A_v(\vec x,t=0) = \fma \, \underbrace{\sqrt{\gamma\, \mathcal{R}
  T_0/\mu}}_{=c_{\text{sound}}}\,\exp{
  \left[-\frac{1}{2}\left(\frac{y}{\beta H_p/2}\right)^{2}\right]}\ .
\label{eq:vamp}
\end{align}
The parameter $\fma$ therefore sets the peak Mach number in the vertical
velocity amplitude. Using the relations \cref{eq:At,eq:Au,eq:Ap} from the theory
described in \cref{appendix:busigw} one finds for the initial conditions
\begin{alignat}{3}
  \vartheta(\vec x,t=0) &= \thse+&&\mathfrak{R}\left\{ -\frac{i}{\omega}\frac{\dd\thse}
    {\dd y}A_v e^{\ii\vec k\cdot \vec x }\right\}, \\
  u(\vec x,t=0) &= &&\mathfrak{R}\left\{ -\frac{k_y}{k_x}
    A_v e^{\ii\vec k\cdot \vec x}\right\}, \\
  v(\vec x,t=0) &= &&\mathfrak{R}\left\{
    A_v e^{\ii\vec k\cdot \vec x}\right\}, \\
  p(\vec x,t=0) &= p_{\mathrm{hse}}+&&\mathfrak{R}\left\{
    -\rho_0\omega \frac{k_y}{k_x^{2}}A_v e^{\ii\vec k\cdot \vec x}\right\},
\label{eq:initial}
\end{alignat}
where $\mathfrak{R}\left\{.\right\}$ denotes the real part of a complex
expression.

In \cref{subsec:packetevo}, the time evolution of an initial amplitude
modulation of the form of \cref{eq:vamp} is derived while considering the IGW
dispersion relation \cref{eq:bdisp} up to second order in $\vec k$. The result
given by \cref{eq:aampxf} shows that the initial profile broadens in time and
moves at a vertical velocity of
\begin{align}
  c_{gy} = \left|\sin\theta\,\cos\theta\,\right|\, \frac{N_0}{|\vec k|}
  \label{eq:gvelsetup}
\end{align}
which corresponds to a vertical Mach number of
\begin{align}
        \gma &= \left|\sin\theta\,\cos\theta\,\right|\, \frac{\sqrt{\gamma
  -1}}{\gamma}\,\frac{\beta}{2\pi} \label{eq:gma} \\
    &\approx \num{0.034} \, \beta.
\end{align}
In \cref{subsec:bigwres} we compare the predicted to the simulated evolution of
the velocity amplitude to assess the accuracy of our numerical schemes. In
order to extract the velocity amplitude function from the simulation,
\citet{miczek2013a} suggests the following procedure:
It is assumed that the horizontal velocity field can be decomposed as
\begin{align}
  u(x,y) = \hat u(y) \, \sin\left( k_x x + \varphi(y) \right).
  \label{eq:ux}
\end{align}
This is fulfilled for the ansatz \cref{eq:2Dwave} and the amplitudes as
defined above. Accordingly, also the evolved initial data should at least
approximately fulfill this decomposition. However, directly comparing this to
the prediction is not straight forward.
To simplify the interpretation, the
square of \cref{eq:ux} is integrated over the full horizontal width
\begin{align}
  \int_{0}^{2\pi/k_x} u(x,y)^2\, \dd x = \frac{\pi}{k_x}\hat u(y)^2.
  \label{eq:igwintegral}
\end{align}
The integral can be calculated numerically using the data from the simulation
and therefore provides the possibility to recover the vertical profile $\hat
u(y)$.

The normalized amplitude
\begin{align}
  r_j =  \frac{\sqrt{\frac{k_x}{\pi}\sum_i u_{i,j}^2\,\Delta x}}
    {\left|\frac{k_y}{k_x}A_v(y=0)\right|}
  \label{eq:rj}
\end{align}
then measures the change of the evolved data relative to the maximum of the
initial data. In \cref{eq:rj}, $i$ and $j$ are the indices in horizontal and
vertical direction, respectively; $\Delta x$ refers to the size of the uniform
grid spacing.
The velocity by which the peak of $r_j$ moves upward is interpreted as the
group velocity in \cref{eq:gvelsetup}.

As discussed in \cref{subsec:nonlin_parameter}, one can define a nonlinearity
parameter,
\begin{align}
\varepsilon = \frac{u}{\omega}k_x,
\label{eq:nonlin2}
\end{align}
where $u$ and $k_x$ denote the horizontal velocity and wave number,
respectively. If $\varepsilon \gtrsim 1$ one expects nonlinear effects to
become dominant. Inserting the corresponding expressions into
\cref{eq:nonlin2} gives
\begin{align}
  \varepsilon_{B}
  & = \sin^2{\theta} \; \frac{\fma}{\gma}
\label{eq:nonlinbus}
\end{align}
for the Boussinesq IGWs.

\subsection{IGW SLH results}\label{subsec:bigwres}

In this section we present simulations of the IGW setup described above
for different parameter settings. For comparison, we show the results for the
low-Mach solver \ausmpup and the classical \roe \citep{roe1981a}.

The possible parameter space is restricted by two conditions:
\begin{enumerate}[(a)]
  \item We require $\varepsilon_B \ll 1$ to stay within the linear regime;
  \item The Boussinesq approximation requires $\beta \ll 1$.
\end{enumerate}
As free parameters we choose $\varepsilon_B=\num{e-2}$ and vary $\gma$. The
values for $\beta$ and $\fma$ are then calculated accordingly. This way, we are
able to assess the capabilities of both schemes for different Mach numbers. The
numerical settings for all of the simulations presented in this section are
listed in \cref{tab:bigw}. In particular we have chosen the time step size of
the implicit time stepping such that the period corresponding to the \bvf is
resolved by \num{20} time steps.

\begin{figure}
  \centering
  \includegraphics[width=\columnwidth]{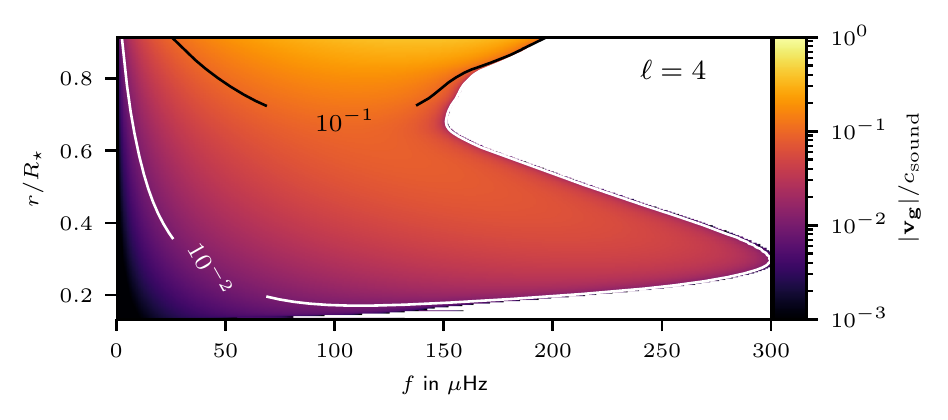}
  \caption{Expected group velocity according to \cref{eq:vgroup}
  expressed in terms of Mach number for values of the stellar model used in
  \cref{sec:results}. Contour lines mark regions of different typical
  Mach numbers. There are no IGWs for frequencies above the \bvf (white
  area). Instead, one expects the excitation of sound waves only which
  have Mach numbers of unity.}
  \label{fig:vgroup_pcolor}
\end{figure}

We perform simulations for vertical group velocities of $\gma=\num{e-4},\
\num{e-3},\ $ and $\num{e-2}$. Whereas condition (b) is well fulfilled for
$\gma=\num{e-4}$, with $\gma=\num{e-2}$, which is closer to the typical
velocity we find in the simulation presented in \cref{sec:results} (see
\cref{fig:vgroup_pcolor}), we have $\beta=\num{3e-1}$ and the stratification
does not strictly follow the Boussinesq approximation anymore.
\citet{press1981a} shows that for the locally Boussinesq but globally anelastic
equations the amplitude scales during the propagation with
$\sqrt{{\rho_0}/{\rho}}$ where $\rho_0$ is the density at the starting point
(cf. \cref{eq:amplification}). We therefore multiply equation \cref{eq:aampxf}
by the correction factor
\begin{align}
  f_\rho = \sqrt{\frac{\rho\left(y-c_{gy}t\right)}{\rho(y)}}
  \label{eq:rhoamp}
\end{align}
to account for the amplification due to varying density.

\begin{figure}
   \centering
  \includegraphics[width=\columnwidth]{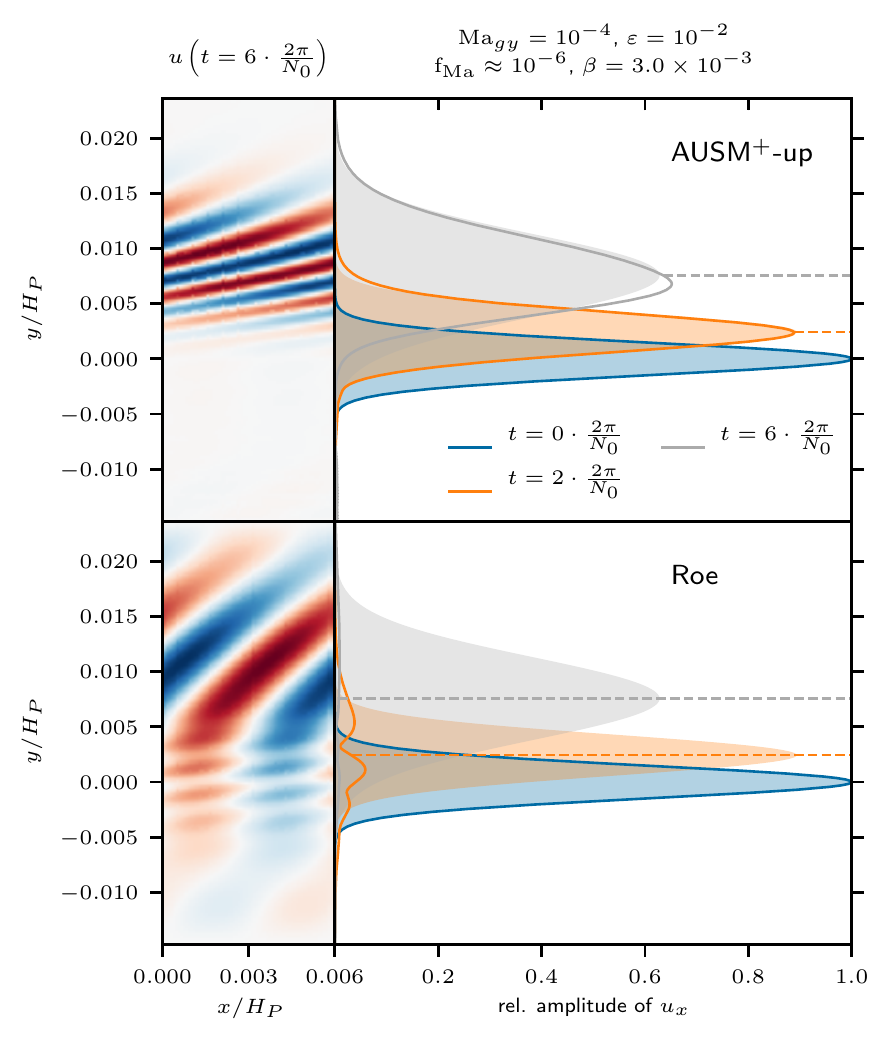}
  \caption{Results for the IGW test setup as described in
  \cref{sec:boussinesq} for the \ausmpup solver \textbf{(upper row)}
  and the classical \roe \textbf{(lower row)}. The parameters for the
  simulation are shown at the top of the plot and described in the
  main text. \textbf{Left column:} Horizontal velocity $u$ in the 2D domain
  at the end of the simulation at $t=\SI{6}{\obvf}$.  Blue color
  corresponds to a positive value, and red color to a negative value
  of the velocity. The scale is adjusted to the maximum amplitude for
  each run.  \textbf{Right column:} Amplitude extracted according to
  \cref{eq:rj} at the beginning of the simulation and at two later
  points in time (solid lines). The shaded areas correspond to the
  predicted shape of the amplitude modulation function according to
  \cref{eq:aampxf,eq:rhoamp}. Dashed horizontal lines mark the position of the
  peak amplitude for the prediction that moves at the group velocity according
  to \cref{eq:gma}.}
  \label{fig:gpack1}
\end{figure}

\begin{figure}
  \centering
  \includegraphics[width=\columnwidth]{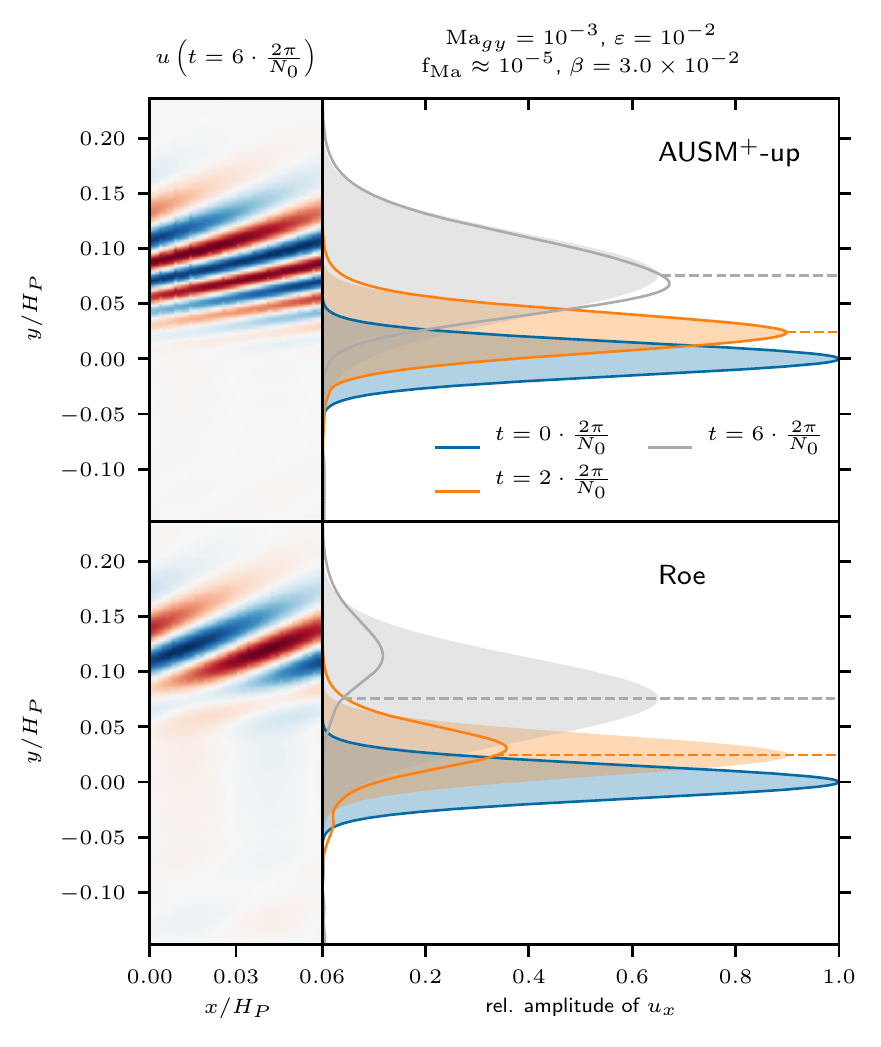}
  \caption{See \cref{fig:gpack1} for a description of the quantities shown.}
  \label{fig:gpack2}
\end{figure}

\begin{figure}
  \centering
  \includegraphics[width=\columnwidth]{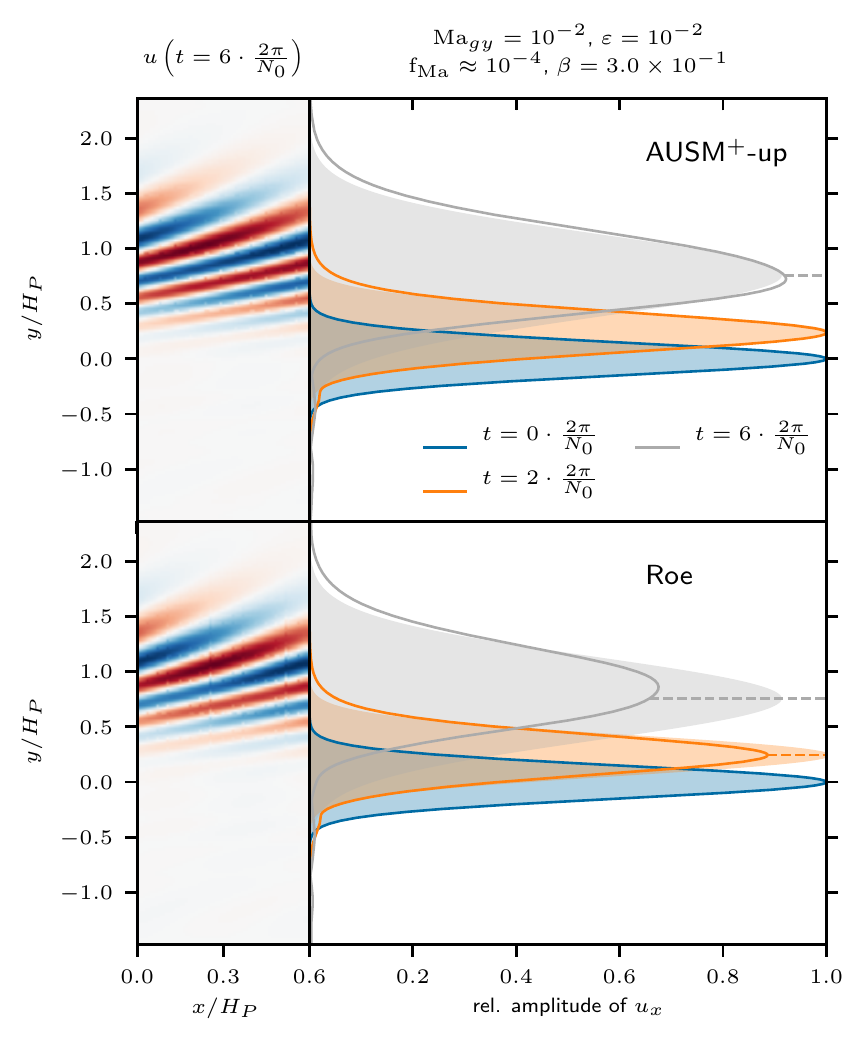}
  \caption{See \cref{fig:gpack1} for a description of the quantities shown.}
  \label{fig:gpack3}
\end{figure}

The results for the \ausmpup and Roe solvers are visualized in
\cref{fig:gpack1,fig:gpack2,fig:gpack3}, respectively. The left columns show
snapshots of the horizontal velocity $u$ at the end of the simulation.  The
right columns compare at three points in time the amplitude and shape of the
vertical velocity distribution as extracted from the simulation using
\cref{eq:rj} with the approximate prediction given by
\cref{eq:aampxf,eq:rhoamp}.

In \cref{fig:gpack1}, we have set $\gma = \num{e-4}$, which corresponds to a
vertical velocity amplitude of $\fma=\num{e-6}$. For the \ausmpup solver the
velocity field in the left column clearly shows the effect of dispersion: waves
of longer vertical wavelengths move faster and therefore appear at larger $y$
values compared to smaller vertical wavelengths. The amplitude function
broadens over time and is compatible with the prediction in terms of width and
peak amplitude. We attribute most of the small deviations from the prediction
to our neglect of third order effects in the dispersion relation, which are
responsible for the amplitude function's skew. In contrast, the \roe heavily
damps the initial amplitude within the first few time steps and it becomes
impossible to determine a unique peak in the remaining velocity field. This
illustrates that the classical \roe fails in the very low Mach regime whereas
\ausmpup still gives excellent results.

We continue by increasing the vertical group velocity to $\gma=\num{e-3}$.
According to \cref{eq:gma} this can only be achieved by increasing $\beta$,
which sets the fraction of the pressure scale height covered by one wavelength,
to $\beta=\num{3.0e-2}$. The results are shown in \cref{fig:gpack2}. The
relative peak amplitudes and shapes for \ausmpup do not change considerably
compared to \cref{fig:gpack1}. The results of the \roe show distinguishable,
but still strongly damped, peaks and their vertical group velocity considerably
disagrees with the theoretical prediction.

The effect of increasing the vertical group velocity further to $\gma =
\num{e-2}$ is depicted in \cref{fig:gpack3}: Again, the shape and peak
amplitudes for \ausmpup are compatible with the prediction. For this setup,
$\beta =\num{3e-1}$ and consequently the density varies noticeably in the
simulation volume. This is reflected by the smaller decrease in the expected
peak amplitudes in \cref{fig:gpack3} compared to the amplitudes shown
\cref{fig:gpack1,fig:gpack2} for which the amplification according to
\cref{eq:rhoamp} is negligible. With the \roe, the results are better as
compared to those obtained at lower vertical group velocities and the
broadening is closer to the prediction. However, significant damping of the
velocity amplitude is still evident.

In summary, the results for the classical \roe, which we present here solely
for comparison, clearly suffer from high damping and show that the wave
packages move at the wrong speed. Our tests therefore confirm the need for
specialized low-Mach solvers in order to treat IGWs in the regime of velocities
below $\mathrm{Ma} \sim \num{e-2}$. A variety of such solvers are readily
available in \slh. A promising example is the \ausmpup solver for which our
tests demonstrate its capability to treat IGW in the low-Mach regime.

To assess the relevance of this finding we estimate the order of magnitude of
the group velocity we expect for the simulation of the real stellar setup
presented in \cref{sec:results}. We take the absolute value of \cref{eq:gvel}
and rewrite it in terms of vertical and horizontal components of the wave
vector. For the polar coordinates used in the 2D simulation, the horizontal
wave number corresponding to the angular degree $\ell$ is given by
\begin{align}
  k_h = \frac{\sqrt{\ell(\ell+1)}}{r}
\end{align}
and the absolute value of the group velocity is
\begin{align}
  |\vec{c_g}(r)| = \frac{r\omega^2}{N\sqrt{\ell(\ell+1)}}\,
  \sqrt{\frac{N^2}{\omega^2}-1}, \label{eq:vgroup}
\end{align}
where $r$ denotes the radial position within the model.
In \cref{fig:vgroup_pcolor} we show the group velocity for $\ell=4$ when
inserting the values from the stellar model used in \cref{sec:results} into
\cref{eq:vgroup}. It illustrates that there are large regions at low
frequencies with Mach numbers around and below \num{e-2}. These regions become
even more extended at higher $\ell$-values. We conclude that low-Mach solvers
are needed to correctly describe the wave field in simulations such as that
presented in \cref{sec:results}, especially in the low-frequency regime
dominated by internal gravity waves.

\section{\texorpdfstring{2D simulation of a \SI{3}{\msol} ZAMS star} {2D 
simulation of a 3Msol ZAMS star}}\label{sec:results}

The previous section demonstrates the capabilities of the \slh code and the
methods implemented therein to propagate IGWs in the low-Mach regime whereas
classical approaches fail. In this section, the \slh code is applied to a real
stellar setup, which encompasses both the generation and propagation of IGWs and
sound waves.

\subsection{Initial model}\label{subsec:initmodel}

For the simulation presented here we use the identical initial 1D model as
\edel. It describes a nonrotating \SI{3}{\msol} star at the ZAMS with a
metallicity of $Z= \num{e-2}$ and an outer radius of $\text{R}_\star =
\SI{1.42e11}{\cm}$.  The model has been calculated with the open-source stellar
evolution code MESA (see \citealp{paxton2019a} for the latest report on code
updates).

The 1D data provided by the MESA model needs to be mapped onto the \slh grid
while accurately fulfilling the equation of hydrostatic equilibrium
\begin{align}
   \nabla P = \vec g \rho.
   \label{eq:hse}
\end{align}
To do so, we reintegrate \cref{eq:hse} while imposing the radial profile of
one thermodynamic quantity from the 1D code (using the 1D density profile for
this would be a simple example). All other thermodynamic quantities then follow
from the equation of state. The particular choice of the imposed quantity
depends on the specific setup. For the case presented here, it is important to
keep the position of the convection zone as close as possible to the 1D input
MESA model. Convective instability is characterized by a negative sign of the
\bvf. In regions without any or with only a small gradient in composition (well
fulfilled in ZAMS stars) it is essentially determined by the sign of
$\nabla-\nabla_{\mathrm{ad}}$. Therefore, we follow the approach of
\citet{edelmann2017a} to integrate \cref{eq:hse} while enforcing
\begin{equation}
  (\nabla-\nabla_{\mathrm{ad}})_{\mathrm{SLH}} =
  (\nabla-\nabla_{\mathrm{ad}})_{\mathrm{MESA}}.
  \label{eq:nabnabad}
\end{equation}
This way the initial spatial extent of the convective zone on the \slh grid
exactly matches the 1D input model. Consequently, other quantities might
deviate from the 1D model.
\begin{figure}
  \centering
  \includegraphics[width=\columnwidth]{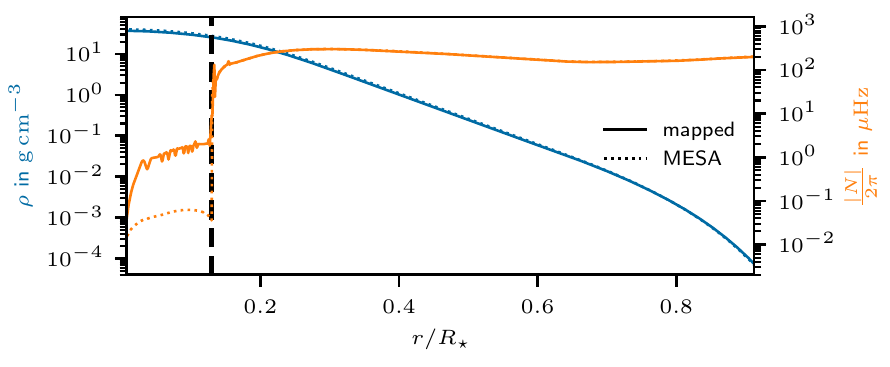}
  \caption{Density (blue) and absolute value of the \bvf (orange). The dotted
  lines correspond to the profiles of the underlying 1D MESA model. The solid
  lines result from our mapping to the \slh grid. The vertical dashed line
  marks the radius below which the \bvf has a negative sign and one therefore
  expects convection to develop. The applied mapping method guarantees that the
  position of the dashed line is identical for the mapped \slh model and the 1D
  MESA model.}
  \label{fig:input}
\end{figure}
In \cref{fig:input} we exemplarily compare the profiles of the \slh model and
the 1D input for density and \bvf. Both quantities are reproduced reasonably
well considering that we cover several orders of magnitudes.There is a
deviation of a factor of ten in the convection zone which we attribute to
differences in the technical details of the equation of state and the
calculations of gradients between MESA and \slh. It is not expected that
enforcing $N^2_{\mathrm{SLH}}=N^2_{\mathrm{MESA}}$ will improve the result as
this will only translate the differences into other quantities.  However, in
the \slh simulation the value of $N^2$ in the convection zone will
self-consistently adjust to a value that corresponds to the equilibrium between
energy input, for example due to nuclear burning, and the convective flux. The small
initial deviation is therefore not relevant.  For the \bvf, the mean
differences in the radiation zone between the 1D MESA model and the mapped
profiles is \SI{0.56}{\percent} with a maximum value of \SI{1.5}{\percent} at
$r=\SI{0.87}{R_\star}$ and we therefore consider the applied method to be
sufficiently accurate.

The underlying 1D MESA model is in thermal equilibrium. The reintegration of
\cref{eq:hse} for hydrostatic equilibrium changes the 1D stratification only
slightly and we expect the mapped \slh stratification in the radiative envelope
to be sufficiently close to the equilibrium model. This, however, is not true for
the convective core where we have to artificially boost the nuclear energy
release (see \cref{subsec:2dslh}). A stratification that is not in thermal
equilibrium will readjust on the thermal-diffusion time scale $\tdiff$. It
can be estimated as (e.g., \citealp{maeder2009a})
\begin{align}
  \tdiff(\Delta x_{\mathrm{diff}}) \sim \frac{(\Delta x_{\mathrm{diff}})^{2}}{K},\quad 
    K = \frac{4\,a\,c_{\mathrm{light}}\,T^3}{3\,\kappa\,\rho^2\,C_{\mathrm{P}}},
    \label{eq:tdiff}
\end{align}
where $\Delta x_{\mathrm{diff}}$ is a typical length scale. The thermal
diffusivity $K$ introduces the radiation constant $a =
\SI{7.57e-15}{\erg\per\cubic\centi\meter\per\kelvin\tothe{4}}$ and the specific
heat at constant pressure $C_\mathrm{P}$ for the ideal gas. The opacity
$\kappa$ is a function of radius and determined by the physical properties of
the gas. However, for the mapped \slh model, we set
$\kappa_\mathrm{\text{\slh}}$ such that we achieve $K_{\text{MESA}} =
K_{\mathrm{\text{\slh}}}$.  Therefore, the value of
$\kappa_\mathrm{\text{\slh}}$ is not fully consistent in a physical sense but
it allows us to stay closer to the stellar values of thermal diffusion.
\begin{figure}
  \includegraphics[width=\columnwidth]{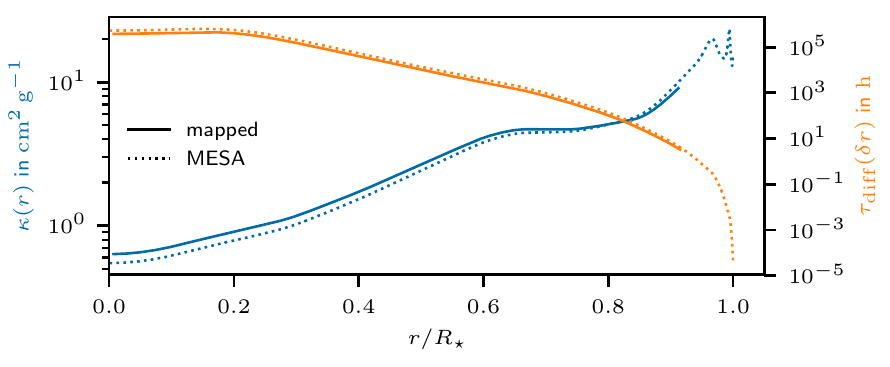}
  \caption{\textbf{Blue lines:} Profiles of the opacity $\kappa$ for
  the underlying 1D MESA model (dotted blue line) and the value
  applied for the mapped 2D \slh model (solid blue line).
  \textbf{Orange lines:} Characteristic diffusion time scale
  according to \cref{eq:tdiff} when assuming the radial spacing
  $\delta r$ of the \slh grid as typical length scale.}
  \label{fig:line_tdiff}
\end{figure}
The blue lines in \cref{fig:line_tdiff} show the profiles of
$\kappa_{\mathrm{\text{\slh}}}$ (solid) and $\kappa_{\text{MESA}}$ (dashed) and
indicate that deviations are reasonably small.

Following \cref{eq:tdiff}, we estimate the thermal-diffusion time scale taking
the radial grid spacing $\delta r$ to be the typical length scale, see
\cref{fig:line_tdiff}. Due to the energy boosting in the convection zone,
deviations from thermal equilibrium are expected to be largest there. From
\cref{fig:line_tdiff} it can be seen that the diffusion time scale is on the
order of \SI{e5}{\hour} and thus two orders of magnitude larger than what is
covered by our 2D simulation (about \SI{e3}{\hour}, see \cref{subsec:2dslh}).
Furthermore, as the grid spacing $\delta r$ is already the smallest possible
scale, our estimate is a lower limit on the time scale. In the outer parts,
the time scale becomes comparable to the simulation time. Thermal flux scales
with $1/r$ in the cylindrical geometry of our 2D simulation (see below for more
details on the numerical setup) and therefore differs from the MESA model where
spherical geometry is assumed. Thus, the thermal flux is not accurately
balanced in our simulation.
However, we do not recalculate the equilibrium state while considering the
correct scaling of the flux in cylindrical geometry as this would lead to large
deviations from the 1D MESA input model. For example, this would likely change
the profile of the \bvf and therefore alter the dynamics of sound and gravity
waves. For our simulations, however, we are interested in waves generated in a
stratification as close as possible to a realistic stellar stratification.

From \cref{fig:line_tdiff} it is clear that slightly imbalanced
initial conditions  most probably only impact the very outer part of our
model during the course of the 2D simulation. To further validate this,  we
performed two 1D simulations. Although they do not include convection and
wave propagation, they reveal the impact of an imbalanced energy flux. The
initial conditions and the numerical settings for both simulations are the same
as for the 2D simulation (see text below and \cref{tab:simulations}) except
that the heat input in the convection zone is turned off. The 1D runs only
differ in geometry, which is cylindrical and spherical, respectively, and cover
the same time span as the 2D simulation. 

In the 1D simulations, the maximum change in temperature occurs close to the
surface and is only $\SI{0.11}{\percent}$. Deviations in the spherical run are
slightly smaller but of the same order as in the cylindrical run. Thus, the
change is most probably due to the Dirichlet boundary conditions for the
temperature which we have chosen for simplicity and that are used for all
radial boundaries in the simulations presented here. To exclude any other cause
than radiative diffusion for the slight change in the background state we also
ran additional 1D simulations where radiative diffusion was disabled. The
stratification remains essentially unchanged in these simulations. For the 1D
simulations with enabled radiative diffusion, we show the resulting radial
profiles of temperature, temperature gradient, and \bvf in \cref{fig:compare1D}
and list the corresponding maximum values in \cref{tab:compare1D}.

While the inconsistent treatment of the thermal flux is
certainly not desirable, the 1D tests show that its impact on this particular
simulation setup is negligible. Nevertheless, for future 2D simulations, we
plan to implement the correct geometrical scaling of the flux and to improve
the flux boundary conditions.

\edel do not need to modify the 1D input MESA model. Therefore,
\cref{fig:input,fig:line_tdiff} serve as a direct comparison between the mapped
model on the \slh grid and their initial data.

\begin{table}[htpb]
  \centering
  \caption{Parameters of the 2D simulation. $\text{R}_\star=\SI{1.42e11}{\cm}$
  denotes the total radius of the underlying 1D model, $\text{L}_\mathrm{MESA}$ the
  stellar luminosity, and $K_\mathrm{MESA}$ the stellar thermal diffusivity as given by MESA.}
  \label{tab:simulations}
  \begin{tabular}{ll}
    \toprule
    boosting             & $L = \SI{e3}{L_\mathrm{MESA}}$          \\
    th.\ diffusivity     & $K = K_\mathrm{MESA}$                   \\
    EoS                  & ideal gas + radiation pressure          \\
    \midrule
    geometry             & polar coordinates                       \\
    radial domain        & \SIrange{0.007}{0.912}{R_\star}         \\
    resolution           & $960 (r) \times 720 (\varphi)$          \\
    boundary conditions  & $r$-direction: solid-wall               \\
                         & $\varphi$-direction: periodic           \\
    time stepping        & ESDIRK23                                \\
    time step size       & $\overline{\delta t} = \SI{8}{\second}$ \\
    time span            & \SI{700}{\hour}                         \\
    reconstruction       & linear, second order in space           \\
    flux function        & \ausmpup \\
    \bottomrule
  \end{tabular}
\end{table}

In \cref{tab:simulations} we list the parameters for the \slh simulation.  We
use 2D polar geometry which corresponds to an infinite cylinder.  \slh
currently does not support the geometry of a 3D sphere with
azimuthal and longitudinal symmetry. Because of the singularity of polar
coordinates at $r=0$, we cannot include the whole core. The minimum radius of
the domain is mainly determined by the decreasing cell sizes in horizontal
direction which affects the possible time step size according to the CFL
criterion (see \cref{subsec:timestepping}). We have chosen $r_{\mathrm{min}} =
\SI{0.007}{R_\star}$ which still allows for reasonably large steps. The upper
boundary was set to $r_\mathrm{max} = \SI{.91}{R_\star}$ which is close to the
value of \edel who set the upper boundary to $\SI{0.9}{R_\star}$.  We apply
solid-wall boundary conditions at the inner and outer boundaries of the
computational domain. They enforce a vanishing velocity perpendicular to the
boundary interface. This prohibits mass flux and sound waves from leaving the
domain.  Periodic boundaries are chosen in the azimuthal direction, which is
appropriate since we cover the full azimuthal range of $2\pi$. The number of
\num{960} radial cells ensures that the smallest pressure scale height (close
to the outer boundary) is still resolved by 16 grid cells. The number of
horizontal grid cells is set such that the cell width at the top of the
convection zone $\delta_w = \delta_\varphi r_{\mathrm{top}}$ roughly matches
the height $\delta_r$ of the cells, where $\delta_\varphi = 2\pi / N_\varphi$
and $\delta_r = (r_{\mathrm{max}} - r_{\mathrm{min}})/N_r$ denote the angular
and radial resolution, respectively.

\subsection{2D SLH results}\label{subsec:2dslh}

For the stellar luminosity as given in the 1D MESA input model, mixing-length
theory (MLT, \citealp[e.g.,][]{kippenhahn2012a}) predicts Mach numbers for the
convective core of around \mbox{$\text{Ma}\sim\num{e-4}$}. Simulations in this regime
are numerically very challenging and the \slh code has several specialized
low-Mach approximate Riemann solvers implemented. However, we have recently
noticed that for Mach numbers considerably below $\num{e-3}$ the convective
flow seems not to be driven by heating but rather by numerical instabilities.
This problem is subject of ongoing investigations but there is no adequate
solution available yet. As a workaround, it was decided to artificially boost
the energy generation in order to increase the convective velocity. From MLT
one finds that the convective velocity $\vconv$ scales with the luminosity $L$
as
\begin{align}
  \vconv \propto L^{1/3}
  \label{eq:vboost}
\end{align}
which has also been confirmed by numerous numerical studies (e.g., \figr7 of
\citealp{cristini2019a} or \figr15 of \citealp{andrassay2019a}).
We boost the stellar luminosity by a factor of \num{e3} which corresponds
to a tenfold increase in velocity compared to the MLT prediction.
We note that this boosting is still a factor \num{e3}
smaller compared to the boosting in the simulations of \edel and
\citet{rogers2013a}. 

\begin{figure}
  \centering
  \includegraphics[width=\columnwidth]{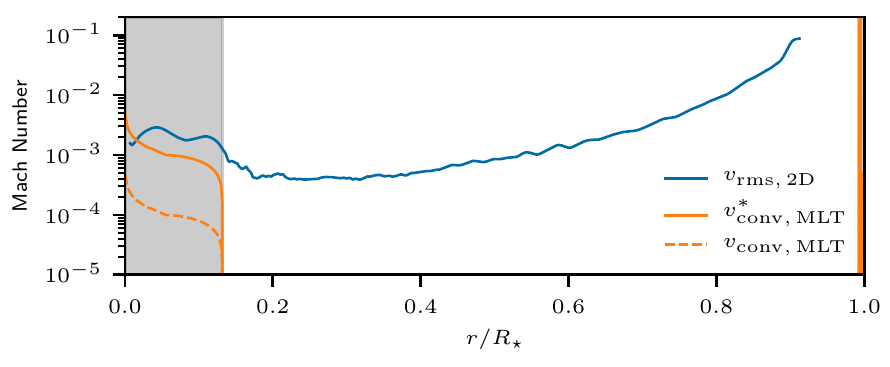}
  \caption{Predicted and simulated Mach numbers of the 2D model. The orange
  lines correspond to the MLT prediction: The dashed profile shows the
  original stellar values, whereas the solid line illustrates the
  velocities scaled by \num{10}, according to \cref{eq:vboost}. The
  blue line shows the 2D results averaged over roughly two convective
  turnover times, starting at $t=\SI{500}{\hour}$. The profile ends at
  $r\approx 0.9 R_\star$ as our domain does not contain the entire
  star model. The gray shaded areas mark the convective core and the small
  region of surface convection in the 1D model.}
  \label{fig:vconv_compare}
\end{figure}

In \cref{fig:vconv_compare} we compare the MLT prediction to the results of our
2D simulation. The gray shaded area at small radii marks the region of the core
that is convectively unstable according to the Schwarzschild criterion. In the
input 1D model, an additional small convection zone near the surface of the
star is present but our 2D model already ends at a smaller radius. We find our
simulations in good agreement with this ``scaled'' MLT prediction.

However, outside of the convective core a velocity field of considerable
amplitude has developed. These velocities are attributed to the excitation and
propagation of IGWs and are of main interest for the work presented here. In
stark contrast to that, no velocities are assumed in the input 1D model. This
illustrates the shortcomings of 1D stellar evolution, where these dynamical
phenomena have to be parametrized \citep[see, e.g.,][]{aerts2019b}.

The speed of sound in our model ranges from \SI{7e7}{\cm\per\second} within the
convection zone to \SI{1e7}{\cm\per\second} at the top of the computational
domain. Accordingly, the Mach number shown in \cref{fig:vconv_compare}
corresponds to typical velocities of \SI{2e5}{\cm\per\second} (convection zone)
to \SI{1.3e6}{\cm\per\second} (near surface). For the mean convective turnover
time in the convection zone $\tconv$ we find
\begin{align}
  \tconv = \frac{2\Delta r_\mathrm{cz}}{v_{\mathrm{rms}}} \approx \SI{40}{\hour}
\end{align}
where $\Delta r_{\mathrm{rc}} = \SI{1.7e10}{\centi\meter}$ is the depth of the
convection zone and $v_{\mathrm{rms}}$ the root-mean-square of the absolute
velocity. The spatial mean has been taken over the convection zone and the
temporal mean includes the entire simulation except for the initial transient
phase, where convection has not yet developed (see also
\cref{fig:line_vel_evo}).  Hence, the \SI{700}{\hour} of fully developed
convection that we follow in our simulation cover roughly \num{17} convective
turnover times. 

\begin{figure*}
  \centering
  \includegraphics{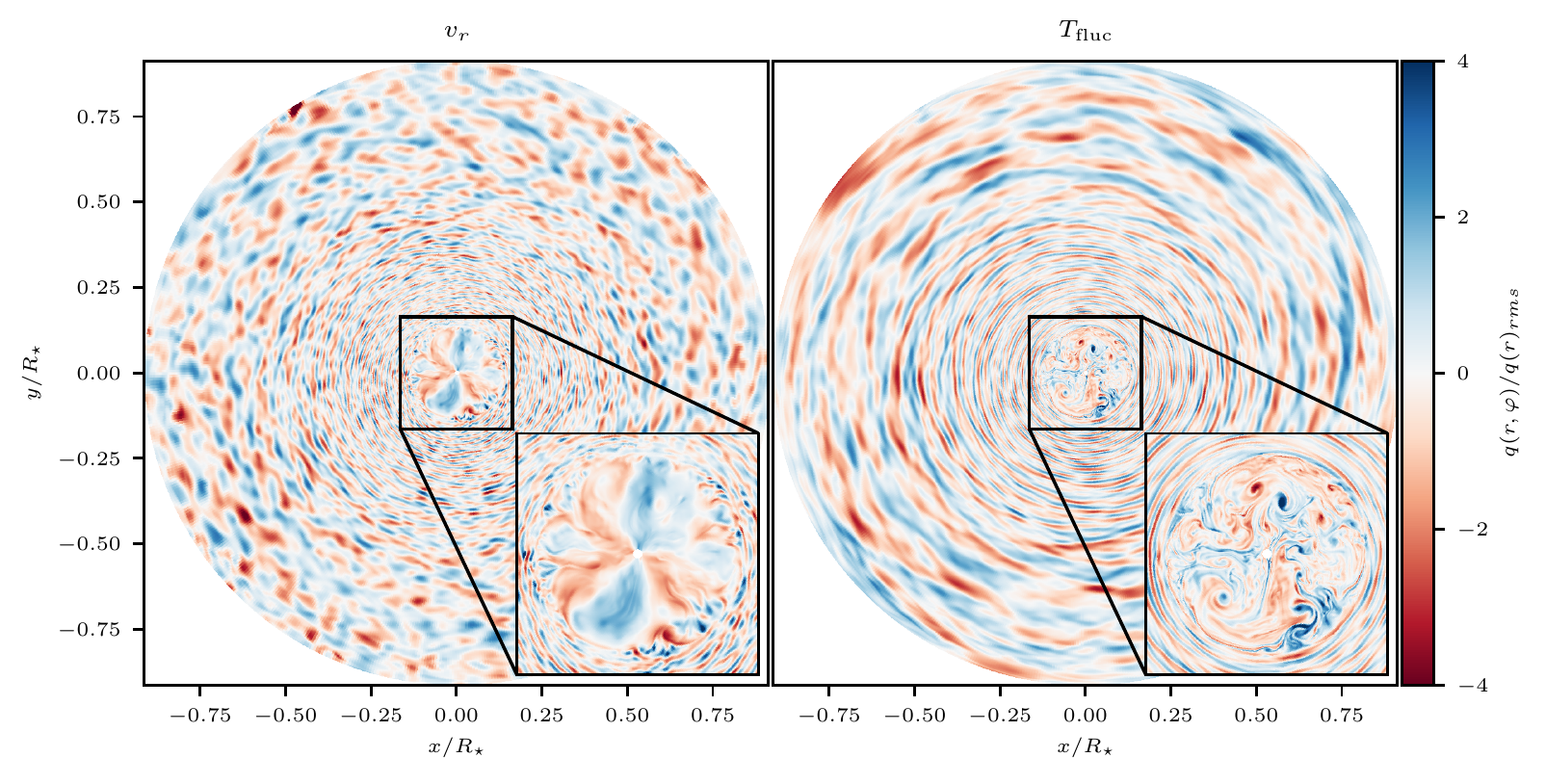}
  \caption{Radial velocity $v_r$ and temperature fluctuations
  $T_\text{fluc}$. As both quantities have larger amplitudes at the outer parts
  of the model (see further \cref{subsec:amplification}) they have been scaled
  with the corresponding horizontal root-mean-square value at each radius $r$
  to ease the visualization. Both panels also show a magnification of the core
  region. The shown snapshot is taken at $t\sim\SI{580}{\hour}$. There is also
  \href{https://zenodo.org/record/3819569}{a movie
  available}.}
  \label{fig:pcolor_velT}
  \centering
  \includegraphics{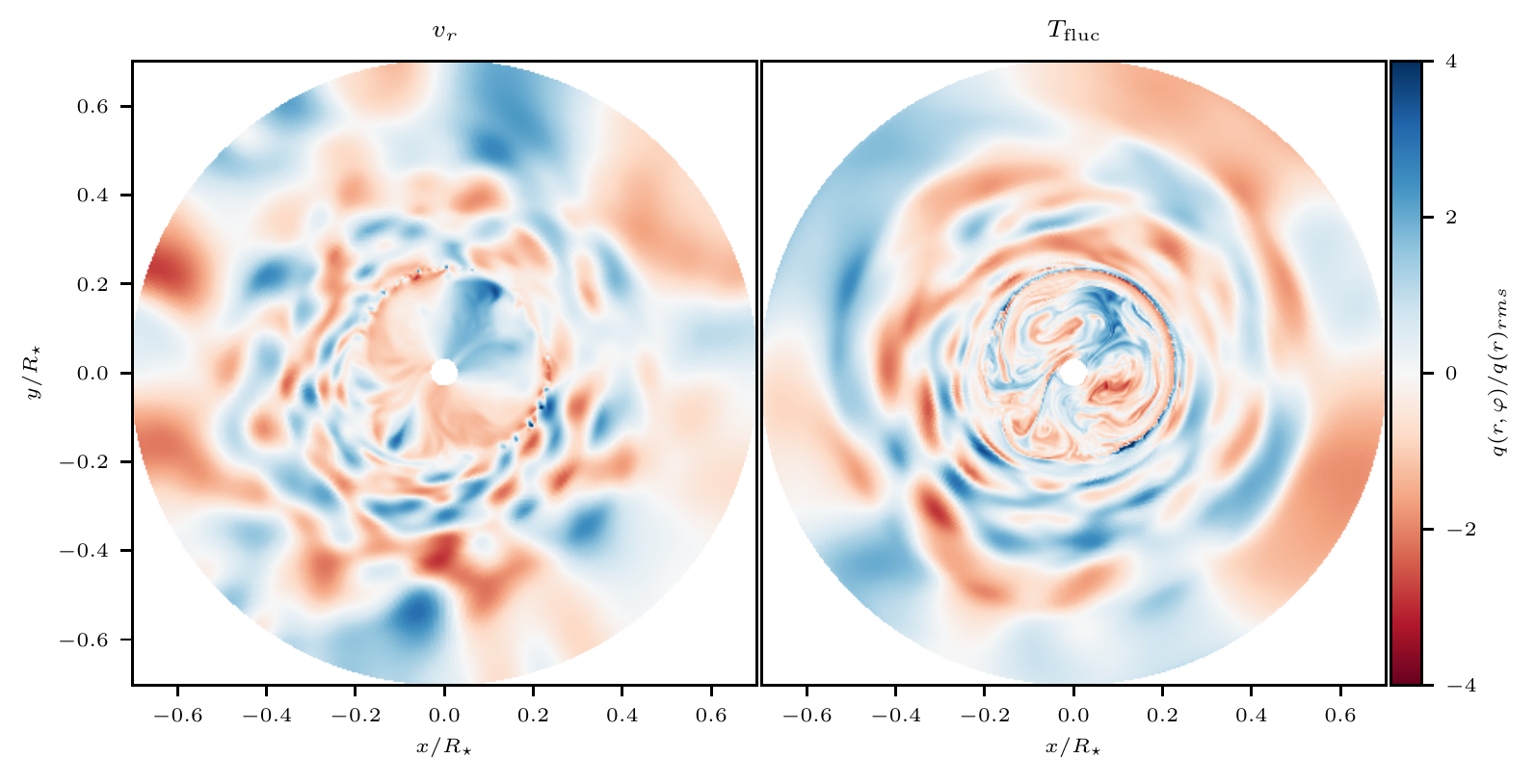}
  \caption{Same quantities as in \cref{fig:pcolor_velT} but here for a 2D
  simulation with a viscosity and thermal diffusivity similar to the
  values used by \edel. We note that the domain for this simulation is
  comparable to \cref{fig:pcolor_velT} but not identical. The radial
  resolution is slightly higher for the viscous simulation and due to the high
  viscosity the energy input is boosted by \num{e6} in order to get convection
  starting.}
  \label{fig:pcolor_velT_visc}
\end{figure*}

The velocity field of the 2D simulation is further illustrated in
\cref{fig:pcolor_velT}. It visualizes radial velocity (left panel) and
temperature fluctuations from the azimuthal average (right panel). Both
quantities are scaled by their horizontal root-mean-square as the amplitudes
vary considerably between the outer and inner part of the model. The relative
amplitudes have maximum values of four and are approximately equal at all radii
for temperature and velocity. In both panels of \cref{fig:pcolor_velT}, the
convective core is filled with a few rather coherent structures which
correspond to convective eddies that induce wave-like motions. These waves form
spiral paths from the point of excitation at the boundary of the convective
core as they travel toward the surface. The spatial structures are smaller
than those observed by \edel (see, e.g., their \figr8).  This is explained by
the much smaller effective viscosity of our simulation and further illustrated
in \cref{fig:pcolor_velT_visc} which shows an \slh simulation of a similar
domain and resolution but with explicit viscosity and thermal diffusivity
comparable to those used by \edel. It is clearly visible how the enhanced
diffusive effects even out small scale structures and only large patterns
remain.

\cref{fig:pcolor_velT,fig:vconv_compare} illustrate that the fundamental
process of generating internal waves, that is a convective core with plumes that
excite waves in the radiative envelope, is present in our 2D simulation. In
order to further validate our results in the subsequent sections, we closely
follow the methods as presented in \edel.

\subsection{Frequency spectra}\label{subsec:spectrum}

The fundamental properties of stellar oscillations can be probed by
investigating their temporal frequency spectrum. For this, we perform a
temporal Fourier transformation (FT) of the entire velocity field in our 2D
simulation.

The transformation is done for both the radial and horizontal velocity. In
order to select waves corresponding to a specific angular degree $\ell$, we
apply a filter prior to the temporal FT\@. For the velocity of each stored
snapshot, a spatial FT in the angle $\varphi$ is taken while using all
available data points on the grid. Subsequently, all amplitudes are set to
zero, except for the one corresponding to the $\ell$ value we want to filter
for. This manipulated spectrum is brought back to real space via an inverse
FT\@. The resulting time sequence of the grid cells of each radial ray is then
multiplied by the Hanning window to reduce leakage effects and serves
as input for the temporal FT\@. To reduce the background noise such that
individual modes appear more clearly, the temporal spectra are taken for
\num{100} evenly distributed radial rays. The squares of the amplitudes of the
resulting Fourier coefficients are then averaged. In principle, the average
could be done for all available angles but we experienced no improvement for a
larger number of rays. Keeping the numbers of rays low is desirable regarding
memory requirements.

The coefficients of the temporal FT are normalized in the same way as in \edel
(see their \eqr{12} and {13}), essentially the amplitudes are divided by the number
of the input data points. This results in coefficients that are independent of
the number of bins and that have the same units as the input quantity.
Furthermore, for this normalization, the amplitude of a peak across one single
frequency bin corresponds to the actual amplitude of that wave in the time
domain which eases the direct interpretation of the spectrum. For our spectra,
the frequency bins have a width of $\delta f = 1 /(\SI{700}{h}) =
\SI{0.4}{\mu\Hz}$. Spectral lines in our spectra typically (except for a few
broad p-modes) span two to four frequency bins and the absolute amplitude is
therefore slightly underestimated when directly read off the spectral plots.
Finally, the amplitudes are multiplied by a factor of \num{2} to 
account for the change in amplitudes in the frequency spectrum due to the
application of the Hanning window (\citealp[see, e.g., Sect. 9.3.9
of][]{brandt2011}).

\begin{figure}[h!]
  \centering
  \includegraphics[width=\columnwidth]{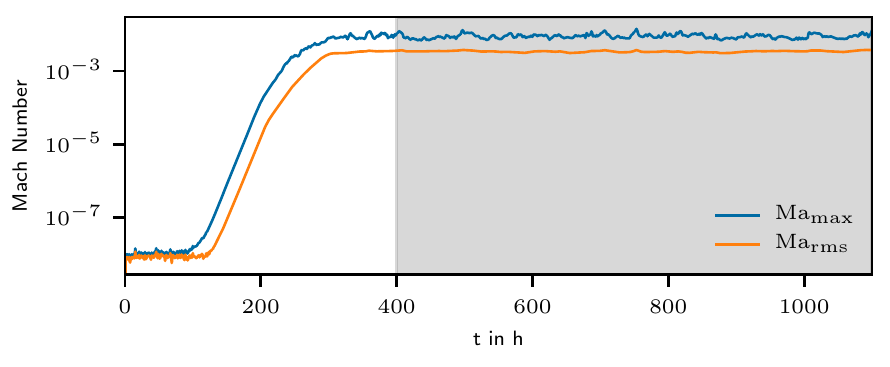}
  \caption{Maximum (blue) and root-mean-square (orange) Mach number
  in the convection zone as a function of time. The gray shaded area marks the time frame
  that has been used to extract the spectra that are presented in
  this paper.}
  \label{fig:line_vel_evo}
\end{figure}

We only use the time interval spanning \SI{700}{\hour} from
\SIrange{400}{1100}{\hour} for the spectral analysis (see the gray shaded area
in \cref{fig:line_vel_evo}) to avoid the initial development of convection in
the core and of the wave field in the envelope.

\begin{figure*}
  \centering
  \includegraphics[width=\textwidth]{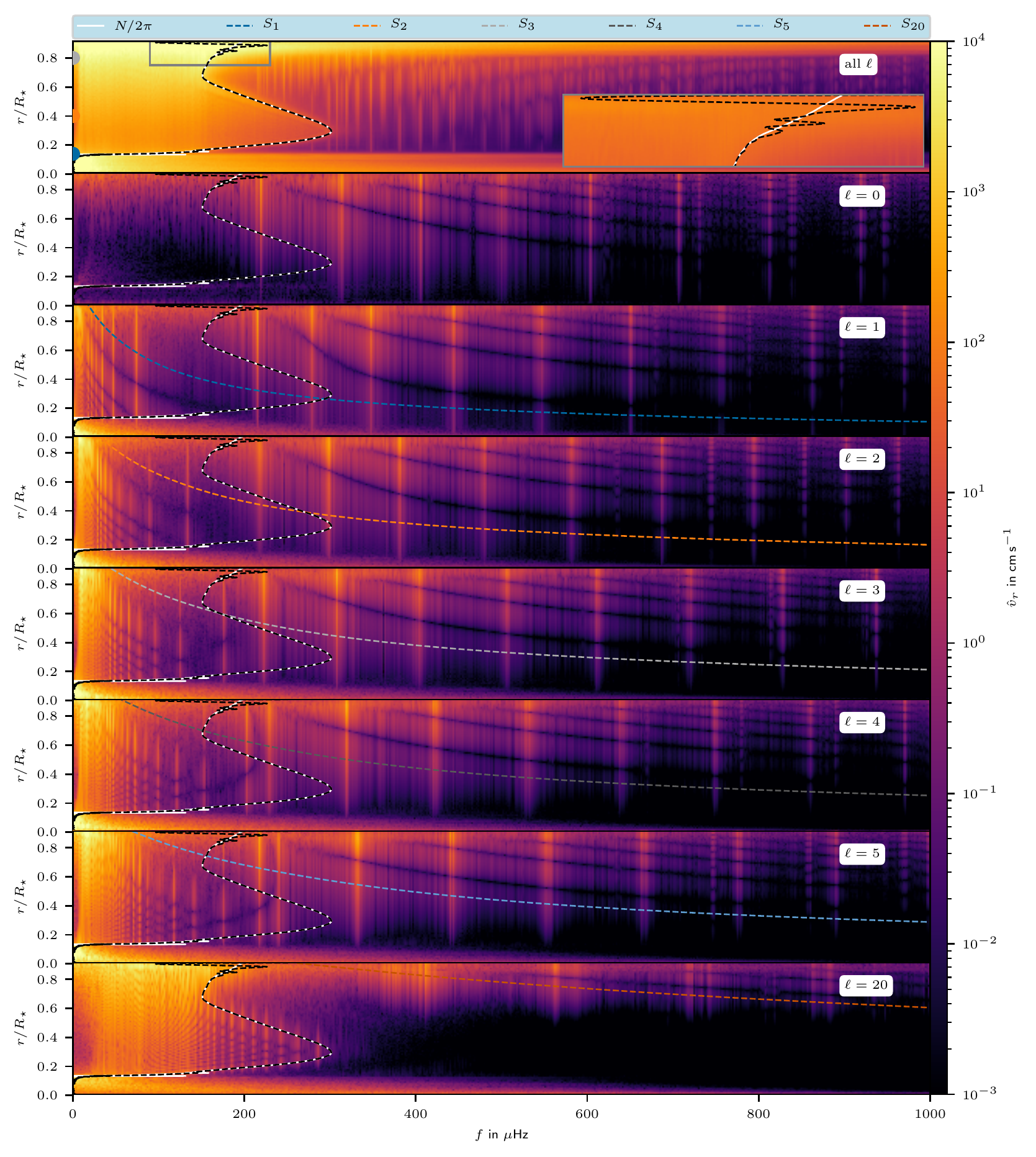}
  \caption{Frequency spectrum of the radial velocity for a time span of
  \SI{700}{\hour}. The normalization is done in a way that the
  amplitude of a narrow line of one frequency bin width corresponds to
  the velocity amplitude of the corresponding  wave in the time domain.
  The data is stored in intervals of \SI{480}{\second}. This allows us
  to capture frequencies up to $f_\text{max} = \SI{e3}{\mu\Hz}$ with a
  resolution of ${\delta f = \SI{0.4}{\mu\Hz}}$. In order to reduce the
  background noise, we show the average of the spectrum of \num{100}
  individual radial rays. The doublets for the modes with $f\gtrsim
  \SI{700}{\mu\Hz}$ are due to aliasing, which we verified with a
  simulation with a very short cadence of outputs. The three colored
  dots in the uppermost panel mark the radii for which the line
  profiles are shown in \cref{fig:line_lallfit}.  The white and black
  lines correspond to the \bvf at start and end of the simulation,
  respectively.  Colored dashed lines correspond to the Lamb
  frequencies of different $\ell$ values according to
  \cref{eq:lambfreq}. In the first row, the uppermost part of the model is
  magnified (gray box) in order to illustrate the change in the \bvf at
  the end of the simulation.  For the magnification, the color scale was
  adapted to increase the visibility of the lines. We show the spectra for the
  horizontal velocity in \cref{fig:pcolor_lall012_h}.}
  \label{fig:pcolor_lall012}
\end{figure*}
\begin{figure*}
  \centering
  \includegraphics[width=\textwidth]{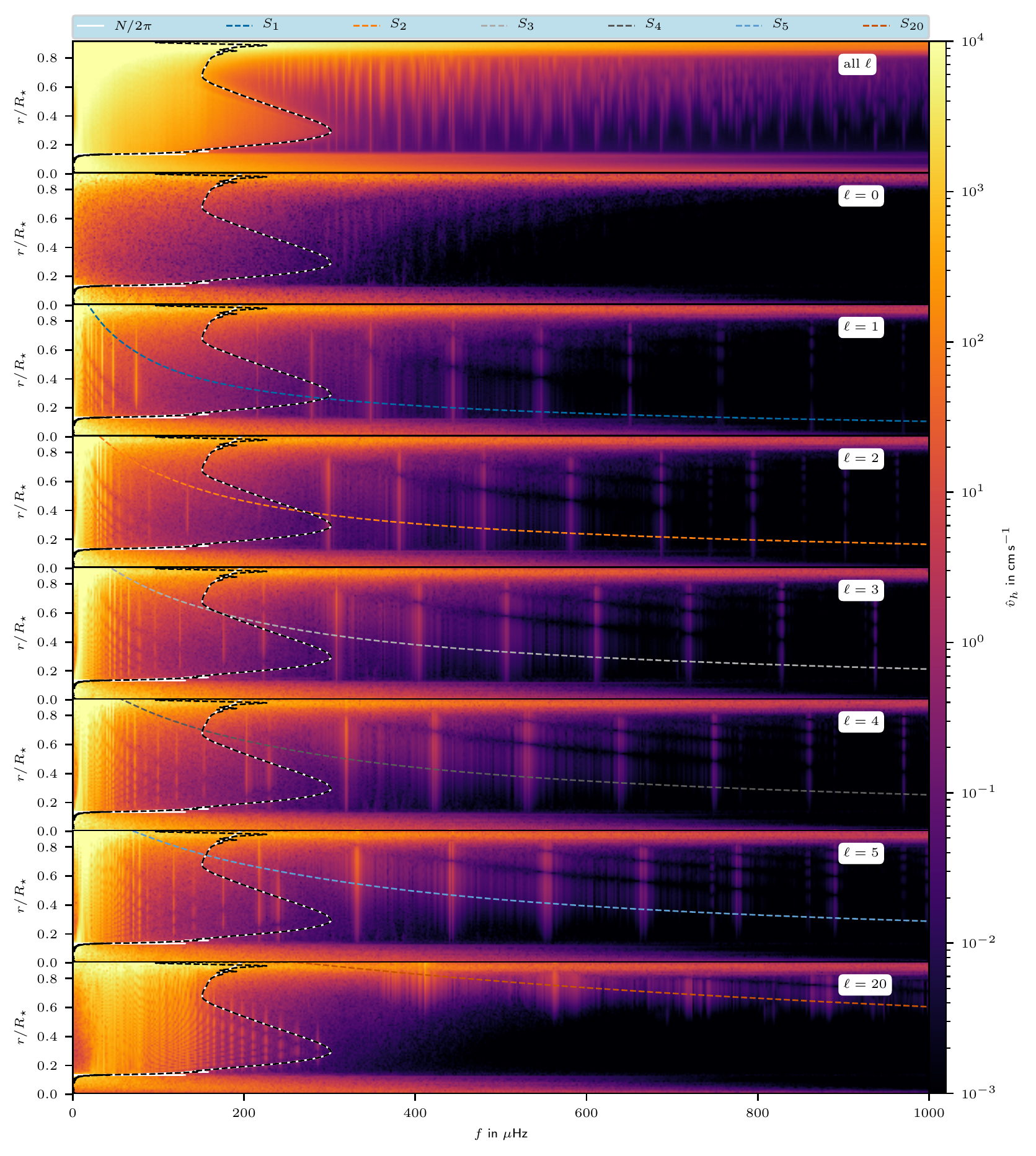}
  \caption{Frequency spectrum for the horizontal velocity, see
  \cref{fig:pcolor_lall012} for details. The p-modes are less prominent as
  their dominant velocity component is radial.}
  \label{fig:pcolor_lall012_h}
\end{figure*}

In \cref{fig:pcolor_lall012} we show the result for the radial velocity, which
is the dominant component for p-modes.  The upper panel shows the spectrum of
the unfiltered velocity, the panels below show the spectra of filtered
velocities for some exemplary values of $\ell$.  This figure is the
compressible counterpart of \figrs{24 and 27} in \edel.  In
\cref{fig:pcolor_lall012_h}, the same quantities are shown for the horizontal
velocity which is the dominant component for g-modes.  In these plots and in
the further course of the paper, a hat denotes quantities that have been
obtained using a temporal FT.

The white line in \cref{fig:pcolor_lall012} indicates the profile of the \bvf
at the start of the simulation, the black dashed line corresponds to the
profile averaged over the last \num{100} hours of the simulation. A magnified
vision of the top of the domain is given for $\ell=\num{0}$ in the gray box. An
oscillatory behavior of the \bvf is apparent at the upper boundary of the
computational domain which is not removed by the time average. Over time,
the oscillations tend to affect a slightly wider range of radii but do
not increase their amplitudes considerably. In the 1D simulation (see
\cref{subsec:initmodel,fig:compare1D,tab:compare1D}), we find that the change
in the \bvf is much smaller and does not develop an oscillatory behavior.
Thus, the oscillations apparent in the outer parts of the 2D
simulation must be due to dynamics that are only captured in multi-dimensional
simulations.

While this behavior could be related for example to numerical effects of our
boundary conditions, grid resolution, or the transport of angular momentum by
propagating IGWs, the exact cause is not fully understood at this point and
still subject of ongoing work. However, the deviation is only within a small
part of the whole model and we do not expect it to have a significant impact on
the general results nor the frequency spectrum of the waves in the envelope
below $R = \SI{0.8}{R_\star}$.  To resolve also the details of the outermost
parts more accurately, a more elaborate outer boundary condition is needed as
well as a higher resolution to resolve the small scale heights accurately
enough.

IGWs can only exist for frequencies below the \bvf (see, e.g., Sect.\ 3.4.2 in
\citealp{aerts2010a}). This property is reflected in the first panel of
\cref{fig:pcolor_lall012} by the fact that the whole possible IGW regime (as
defined by $f<N/2\pi$) shows a significant amplitude which then suddenly drops
for $f>N/2\pi$. This is in qualitative agreement with \figr27 of \edel and a
clear indicator of the excitation of IGWs in the 2D \slh simulation. An
increasing number of radial nodes (or, equivalently, a decrease in the radial
wavelength) for lower frequencies is another general property of IGWs that is
confirmed by the spectra in \cref{fig:pcolor_lall012}.

These features have already been seen and described in \edel. However, our
simulation also shows additional signals. For $\ell=\num{0}$ (second panel in
\cref{fig:pcolor_lall012}) no IGWs are excited because they cannot be purely
radial. Yet, distinct standing modes reaching maximum velocity near the stellar
surface are visible at frequencies larger than the \bvf. Therefore, they must
be signals of excited p-modes.

The colored dashed lines in the panels of \cref{fig:pcolor_lall012} for
$\ell>0$ correspond to the respective Lamb frequencies
\begin{align}
  S_\ell = \frac{\ell(\ell+1)\, c_{\text{sound}}^2}{r^2},
  \label{eq:lambfreq}
\end{align}
\citep[e.g., Sect. 3.3.2 of][]{aerts2010a}. They are derived for spherical
geometry using spherical harmonics and might not strictly hold in our 2D
geometry. However, we still use them here as an estimate to characterize the
general behavior of internal waves. A p-mode of angular degree $\ell$ and
frequency $f_p$ can only exist if the frequency fulfills the conditions
\begin{align}
  f_p > \frac{N}{2\pi} \text{ and } f_p > \frac{S_\ell}{2\pi}. 
\label{eq:pmode}
\end{align}
The corresponding conditions for g-modes are
\begin{align}
  f_g < \frac{N}{2\pi} \text{ and } f_g<\frac{S_\ell}{2\pi},
\end{align}
(see Sect.~3.4 of \citealp{aerts2010a} for further details). In regions where
the frequency does not fulfill these relations, p- and g-modes are
evanescent. This is most easily seen for p-modes in the spectrum
for $\ell=\num{5}$ and for g-modes in the spectrum for $\ell=\num{20}$ in
\cref{fig:pcolor_lall012}.

The occurrence of p-modes is one of the key differences between fully
compressible codes and the anelastic approaches. In the latter, p-modes cannot
occur because sound waves are not included in the equations. To bring the
p-modes of our 2D simulation into context with observations, we compare them to
those of the $\beta\,$Cep stars presented in \citet{aerts2003a}. They find
frequencies at low $\ell$ typically around $f\approx \SI{6}{d^{-1}} \approx
\SI{70}{\mu\Hz}$.  For the particular star \textbf{$\omega^{1}\,\text{Sco}$}
observations indicate $f=\SI{15}{d^{-1}}\approx \SI{174}{\mu\Hz}$ at $\ell =
9$.  Of course, the eigenfrequencies of a real star depend on its stellar
parameters and excitation mechanism, and $\beta\,$Cep stars have masses
typically in the range of \SIrange{8}{20}{\msol}. Thus, the \bvf{}s which set
the minimum p-mode frequencies (see \cref{eq:pmode}) are smaller compared to
our \SI{3}{\msol} model. We find amplitudes associated with p-modes starting at
around \SI{300}{\mu\Hz}. Although our model is not directly comparable to the
observed $\beta\,$Cep stars, this indicates that the waves, which are excited
in our simulations by the convective core, are compatible with those observed
in real stars of similar mass. We note, however, that this cannot be seen as a
proof for the correctness of the model or the underlying excitation mechanism,
especially since p~-modes are not typically observed in main-sequence
3~M$_{\odot}$ stars \citep{aerts2010a}.

For radii deep inside the convection zone ($r\lesssim \SI{.14}{R_\star}$)
distinct modes are less pronounced in
\cref{fig:pcolor_lall012,fig:pcolor_lall012_h} (g-modes generally cannot exist
there  and in our case the Lamb frequencies prohibit also p-modes) but instead
amplitudes in a wide range of frequencies appear, a consequence of the
turbulent convection in the core. This is most easily seen for larger $\ell$
values in the spectrum of the horizontal velocity. On the other hand, also at
the largest radii, a band of high amplitudes is visible in the uppermost panel
which extends over all frequencies. This is due to the development of a shear
flow toward the end of the simulation. As it is located very close to the
outer boundary, we cannot determine beyond doubt whether the shear flow is
caused by the boundary or by deposition of the IGW's angular momentum due to
nonlinear effects. This needs to be examined in more detail in future
simulations.

\begin{figure*}
  \centering
  \includegraphics[width=\textwidth]{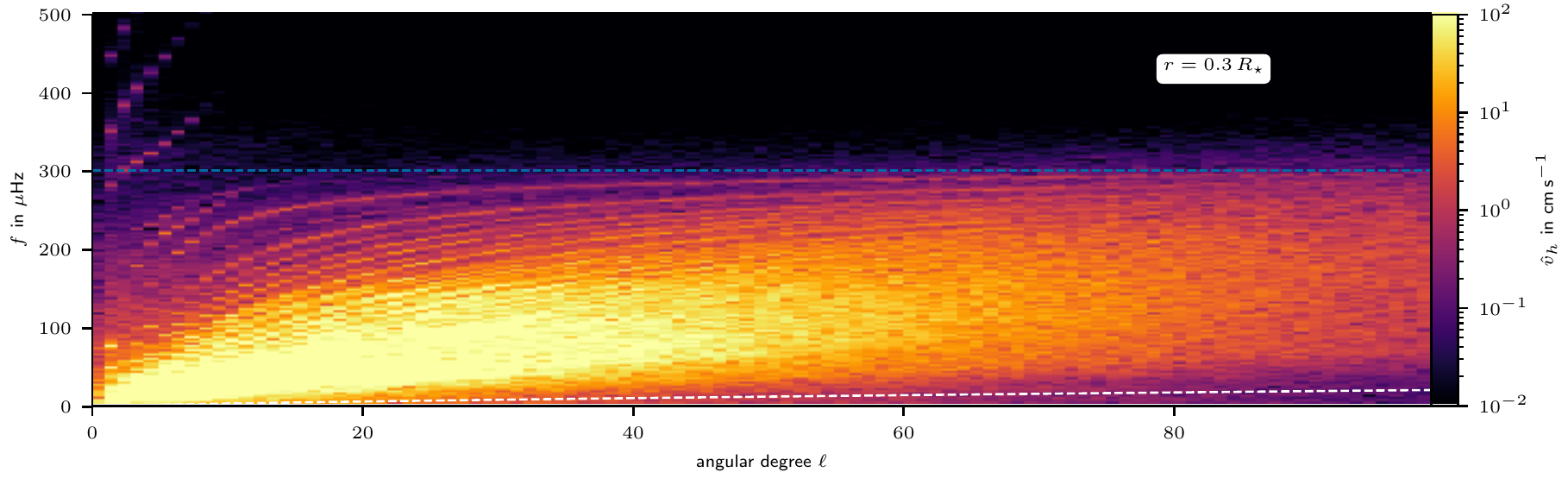}
  \caption{Color map of amplitudes at different frequencies as a function
  of the angular degree $\ell$ at a fixed radius of $R=\SI{0.3}{R_\star}$. The blue
  dashed line corresponds to the \bvf at this radius. The white dashed line
  indicates the cutoff frequency $\omega_c$ below which waves are expected to
  lose too much kinetic energy due to damping to form standing modes.}
  \label{fig:pcolor_lfplane_3}
\end{figure*}

Another way of illustrating the emerging spectra of waves is given in
\cref{fig:pcolor_lfplane_3} (this is similar to \figr22 in
\citealp{herwig2006a} or \figr2 in \citealp{alvan2015a}). Here, the  spectrum
at a fixed radius $r=0.3\,R_\star$ is shown for the first \num{100}
$\ell$-values. The horizontal blue line indicates the \bvf. One clearly
observes distinct modes for small $\ell$ and decreasing mode spacing for
increasing $\ell$. The amplitudes drop at the \bvf as expected for gravity
waves. \citet{alvan2015a} estimate a cutoff frequency $\omega_c$ below which
waves have lost a considerable amount of the kinetic energy due to radiative
diffusion (see their \eqr26). The corresponding profile is shown as dashed
white line. They argue that a traveling wave without sufficient energy cannot
reach the turning point, travel back and interfere with another progressive
wave to form a standing mode.  Their corresponding cutoff is rather steep and
reaches high frequencies, indicating a wide region where they expect
progressive waves. This is similar to \figr5 of \citet{rogers2013a}. In
contrast, for the simulation presented here the profile of $\omega_c$ is much
flatter and stays at much smaller frequencies.  It is clear that the excitation
of gravity waves from core convection produces an entire spectrum of waves
spanning a broad range in frequency.

\begin{figure}
  \centering
  \includegraphics[width=\columnwidth]{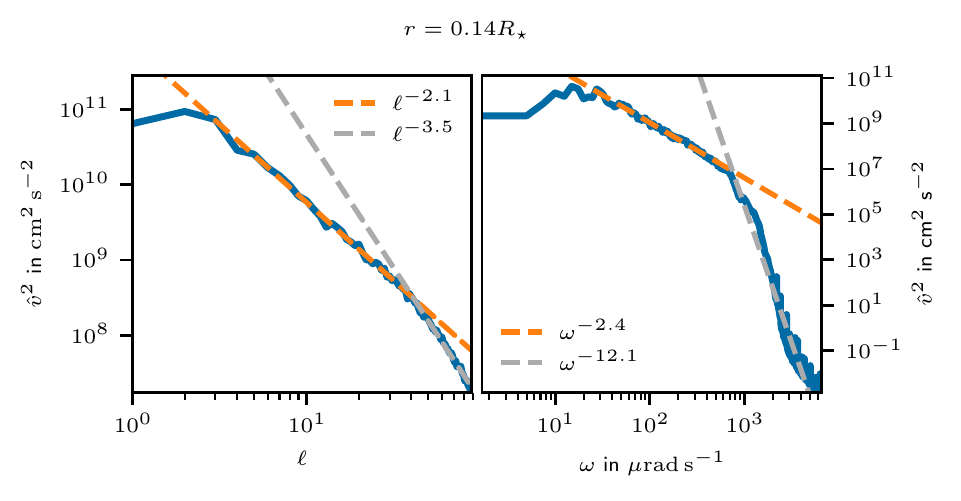}
  \caption{Kinetic energy as a function of angular degree $\ell$ (left panel)
  and angular frequency~$\omega$ (right panel) at the top of the convection
  zone. Dashed lines correspond to power-law fits. This figure
  is similar to the second and third column in \figr6 of
  \citet{rogers2013a}.}
  \label{fig:energy_fit}
\end{figure}

In \cref{fig:energy_fit} we plot the kinetic energies
\begin{align}
  v^2(\ell) = \sum_f \hat v_{\ell,f}^2,\quad
  v^2(\omega) = \sum_\ell \hat v_{\ell,\omega}^2
\end{align}
at the top of the convection zone. Here, $\hat v_{\ell,\omega}$ is the velocity
at angular degree $\ell$ and frequency $\omega$ which results from the spatial
and temporal FT\@. The sum over all degrees $\ell$ is terminated at
$\ell=\num{100}$.  The spectra are fitted by power-laws, as proposed in
\citet{rogers2013a}. Their 2D simulation of a nonrotating model gives similar
values for $v^2(\ell)$, shown in our left panel. Also the position of the
transition between the two regimes is comparable. However, our profiles for
$v^2(\omega)$ (right panel) are much steeper (\citealp{rogers2013a} find
$\omega^{-1.2}$ and $\omega^{-4.8}$).  In the 2D simulation presented here the
transition between the two regimes is at much higher frequencies.

\subsection{Dispersion relations}\label{subsec:disperison}

So far we have presented clear indications of the excitation of IGWs and
pressure waves. To further confirm this, we show in the following section that
the corresponding dispersion relations are fulfilled for both g- and p-modes.

\subsubsection{Dispersion for gravity waves}\label{subsubsec:gdisp}

From theory, one finds a dispersion relation for IGWs of the form (e.g., 
\citealp[Sect. 3.1.4]{aerts2010a} or \citealp[Sect. 3.3.3]{sutherland2010})
\begin{align}
   \frac{\omega}{N} = \frac{k_h}{\left|\vec k\right|}\quad
      \text{with } \left|\vec k \right| = \sqrt{k_h^{2} + k_r^{2}},
   \label{eq:dispersion_gm}
\end{align}
where $\omega=2\pi f$ is the angular frequency of the gravity wave and $k_h,
k_r$ are the horizontal and radial wave numbers. We note that this expression
is derived under the assumption of a spatially constant value for the \bvf.
From \cref{fig:input} it is clear that this is not the case for the stellar
model presented here.  Therefore, \cref{eq:dispersion_gm} is expected to hold
only for wavelengths that are short compared to the scale height of the \bvf. 

To estimate how close our simulation follows \cref{eq:dispersion_gm}, we apply
the method of \edel which is briefly summarized in the following:

The angular frequency $\omega$ and \bvf are readily available quantities.
Because of the spatial FT filtering for specific $\ell$, we know that $k_h^2 =
\ell(\ell+1)/r^2$. The only quantity to determine in \cref{eq:dispersion_gm} is
therefore the vertical wave number $k_r = 2\pi /\lambda_r$ or equivalently the
radial wavelength $\lambda_r$. To determine the wavelength, the positions of
peaks in the amplitude for the FT of the radial velocity are identified for all
available frequencies. The distance between adjacent peaks is interpreted as
one half of the wavelength at the radial position at the midpoint between the
two peaks. We determine the wavelengths starting just above the convection zone
until the upper boundary of the model for all frequencies and interpolate in
radius afterward. This gives for each frequency $f$ the wavelength
$\lambda_{r, f}(r)$. From that, we obtain $k_r$ and are finally able to
evaluate \cref{eq:dispersion_gm} in the entire radius-frequency plane for a
specific order $\ell$.
\begin{figure*}
  \centering
  \includegraphics[width=\textwidth]{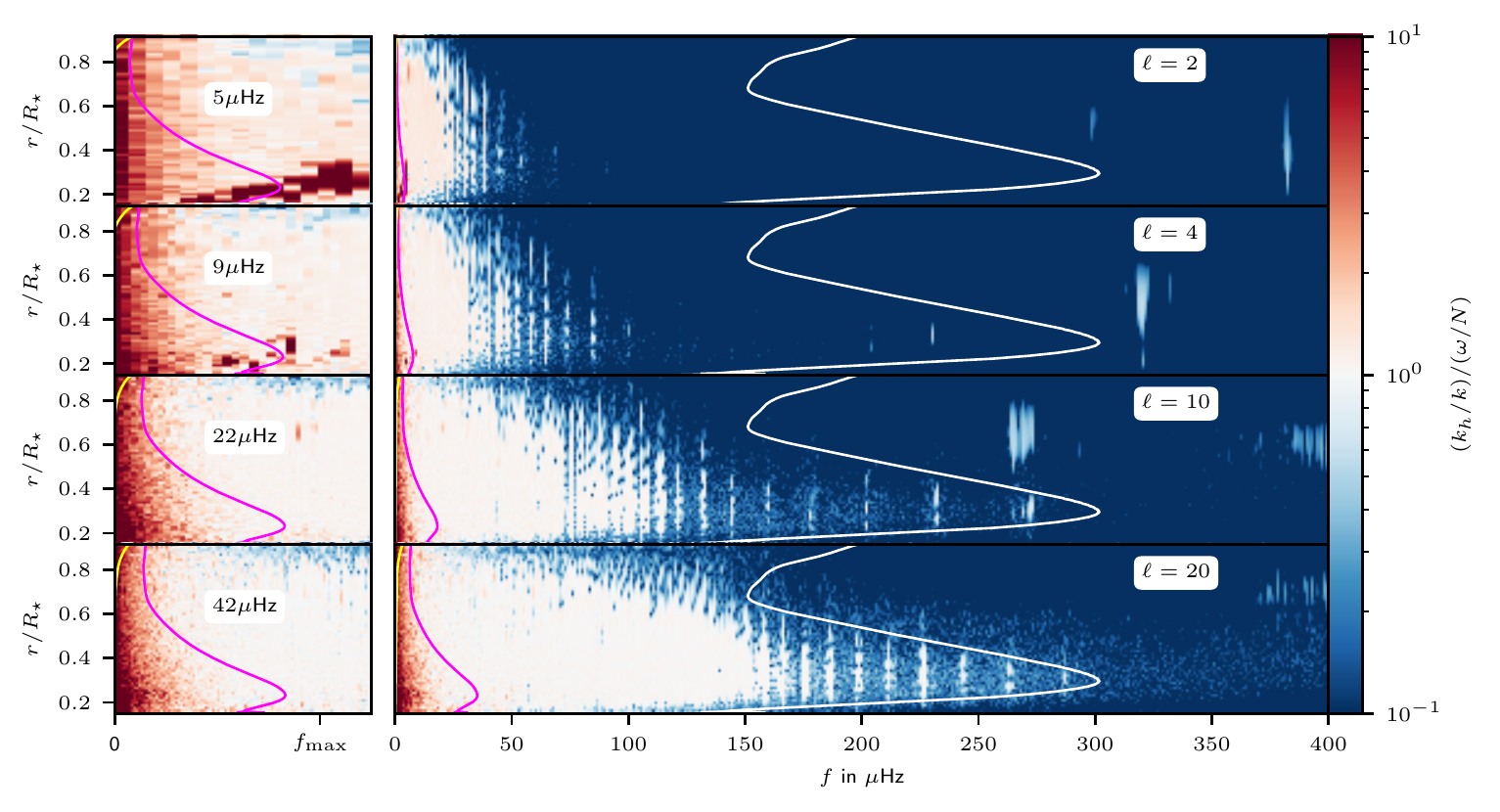}
  \caption{Ratio of the left and right side of the dispersion relation
  \cref{eq:dispersion_gm}.The method to extract $k_h/k$ from the
  simulation is described in the text. A ratio of unity corresponds to
  a perfect match and is reflected by white regions. A red color
  indicates a too large value of $k_h/k$ and correspondingly the ratio
  is too small in blue regions. The white line is the \bvf. For
  frequencies below the yellow line, IGWs are expected to be damped by
  thermal diffusion. Below the magenta line, the radial wavelength is
  resolved by less than 10 grid cells. The left column zooms into the
  low frequency region of the corresponding full plane on the right.
  The frequency given in the white boxes denotes the value of $f_\mathrm{max}$
  which sets the scale of the corresponding x-axis.}
  \label{fig:pcolor_dispersion_gm}
\end{figure*}

To visualize the results, the ratio of the left and right side of
\cref{eq:dispersion_gm} is plotted in \cref{fig:pcolor_dispersion_gm}
exemplarily for $\ell=\text{\numlist{2;4;10;20}}$.  Regions in which the
dispersion relation is fulfilled, that is the ratio is unity, appear white. This
reproduces \figr28 of \edel.  A magenta line denotes the frequency below which
one wavelength is resolved by less than $10$ grid points: For a given radial
grid spacing $\delta_r$, the corresponding wave number is $k_{r,\ n=10} =
{2\pi}/(10\, \delta r)$ such that this frequency is defined as 
\begin{align}
  f_{n=10} = \frac{N}{2\pi} \, \frac{k_h}{\sqrt{k_h^2 + k_{r,\ n=10}^2}}.
  \label{eq:resolved}
\end{align}
The yellow line in \cref{fig:pcolor_dispersion_gm} marks the frequency below
which one assumes the waves to be dissipated by thermal diffusion. The estimate
is based on the equality of the diffusion length and the wavelength at the
critical frequency as given by equation\footnote{Their equation is missing a
factor $\sqrt{N^2/\omega^2 -1}$. It does not change the result qualitatively
but is included here for consistency.}~27 of \edel. In comparison to \edel, the
effect of dissipation is greatly reduced in the \slh simulation. In our case,
we are mainly limited by the radial resolution when going to higher
$\ell$-values. We see agreement with the dispersion relation of linear internal
gravity waves everywhere where such agreement can be expected: for modes that
are well resolved in space but with radial wavelengths short enough that $N^2$
as well as the radius can be considered approximately constant over a single
wavelength. This is the case for low-frequency IGW and gets worse for
increasing radial wavelengths, corresponding to increasing frequencies.  In contrast to
\edel, our results are mainly affected by resolution effects rather than
damping, yet broad spectra of standing modes are clearly excited in both
numerical setups. For a better visibility, the low frequency regions are
magnified in the left column of \cref{fig:pcolor_dispersion_gm}. At
frequencies below \SI{5}{\mu\Hz}, it is difficult to determine if the
dispersion relation is fulfilled because the quality of the FT deteriorates as
the number of wave periods that fit into the time window of our analysis
decreases. However, we are able to reach much smaller frequencies than the
anelastic simulations and provide reliable results for wave frequencies above
\SI{5}{\mu\Hz}.

\begin{figure}
  \centering
  \includegraphics[width=\columnwidth]{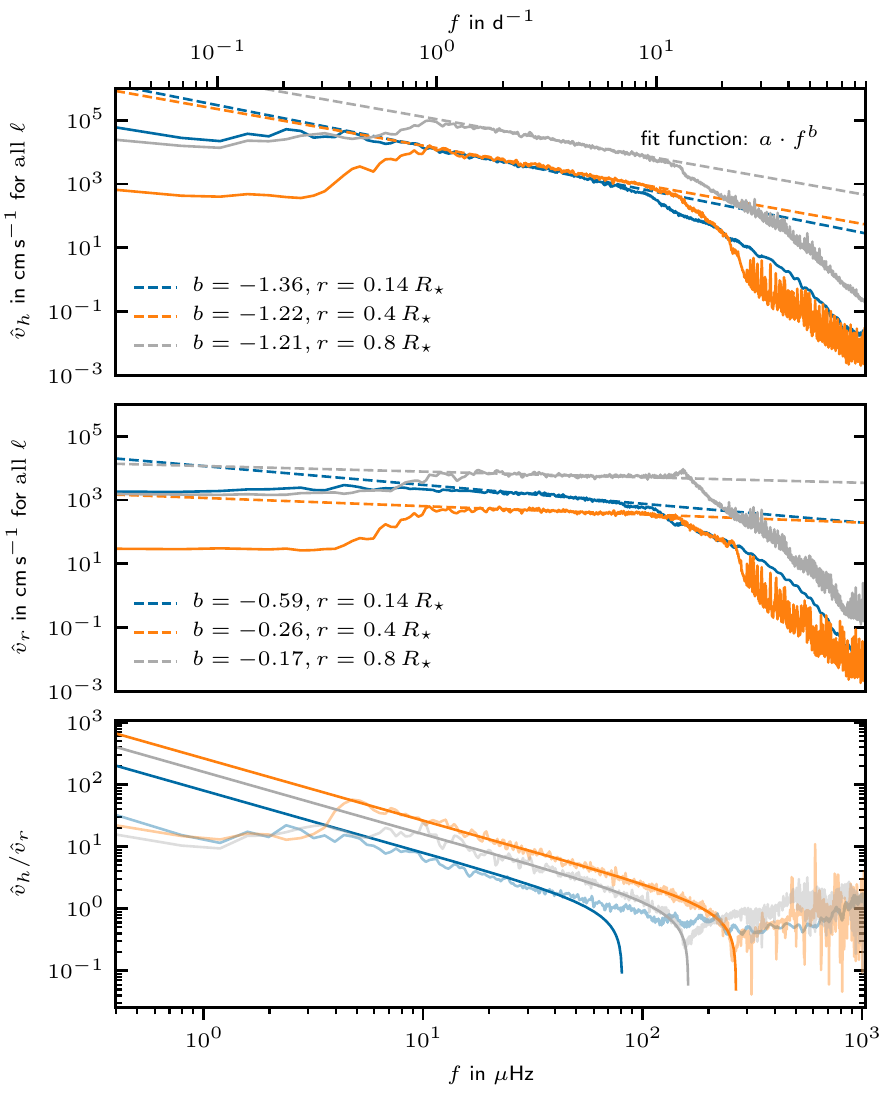}
  \caption{Line plot representations of the
  spectra for horizontal and vertical velocity at three different radii
  \SIlist{0.14;0.4;0.8}{R_\star} are shown in the \textbf{upper two panels}.
  The dashed lines correspond to the power-law fit as denoted in the labels.
  The \textbf{lowest panel} shows the ratio of the velocity components at each
  radius (transparent line) and compares them to the prediction according to
  \cref{eq:vratio} (solid lines).}
  \label{fig:line_lallfit}
\end{figure}

Another way of testing the dispersion relation is to measure the inclination of
the IGW crests with respect to the radial direction. As the fluid velocity in
an internal wave is parallel to the wave crest, we can obtain the inclination
by measuring the ratio $v_h/v_r$ in our simulation. Because the wave vector
$\vec k$ is perpendicular to the crests we have 
\begin{align}
  \begin{pmatrix} v_{r} \\ v_{h} \end{pmatrix}
  \parallel \begin{pmatrix}
  -k_h \\ k_r \end{pmatrix}, \text{ where }  \begin{pmatrix}
  -k_h \\ k_r \end{pmatrix} \cdot \vec k = \vec 0.
\end{align}
It follows that
\begin{align}
  \left|\frac{v_h}{v_r}\right| = \left|\frac{k_r}{k_h}\right| 
\end{align}
and with the aid of \cref{eq:dispersion_gm} we find the expression
\begin{align}
  \left|\frac{v_h}{v_r}\right| = \frac{\sqrt{N^2-\omega^2}}{\omega},
  \label{eq:vratio}
\end{align}
in which the left-hand side can be obtained from the simulation
and compared to the theoretical prediction on the right-hand side.

In the top two panels of \cref{fig:line_lallfit} we show a line plot
representation for the spectra of $v_r$ and $v_h$ at three different radial
positions. They are marked by colored dots in the upper panel of
\cref{fig:pcolor_lall012}. For radii well above the convection zone, the
amplitudes rapidly decrease for $f>N/2\pi$. This is expected for a signal
mostly made up of IGWs which cannot propagate in this frequency range. Above
frequencies of $N/2\pi$ we can see clearly isolated peaks corresponding to
p-modes of various angular degrees. The p-mode amplitudes are at least one
order of magnitude smaller than the IGW signal across a broad range of
frequencies. For the radial velocity, the spectrum is almost flat with a subtle
decrease toward higher frequencies. The spectrum for the horizontal velocity
shows a decrease already within the IGW regime.  Similar to the findings of
\edel, these integrated spectra composed of all angular degrees do not show
distinct peaks. However, our results do not have the steep drop in amplitude at
very low frequencies as seen in their \figr{23}, which can be attributed to the
lower viscosity and thermal diffusivity in our simulation.

In the lowest panel of \cref{fig:line_lallfit} we show the ratio of the
velocity components as depicted in the two panels above (semitransparent
lines). The solid lines represent the prediction given by \cref{eq:vratio}.
There is an excellent agreement for all three radial positions for frequencies
below the \bvf. The line for $0.14\,R_\star$ is just beyond the boundary of the
convective core and therefore does not follow the steep drop in the ratio as
convective motions impact the data. Only for the smallest frequencies the
simulation deviates as a result of damping effects and a lack of independent
data points for the temporal FT\@. This result is another strong indication
that IGWs are correctly represented in the 2D simulation. 

Furthermore, we find the results of our simulations to be comparable to
observations of late-B SPB stars. \citet{decat2002b} report the ratio $v_h/v_r$
for several SPB stars which are typical g-mode pulsators (this corresponds to
the \textit{K}-value in their \figr17).  The value ranges from
\numrange{10}{100}. This observed range is similar to the values shown in the
lower panel of \cref{fig:line_lallfit} for frequencies to the left of the dip,
that is below the \bvf. For p-mode pulsators \citet{decat2002a} (see also
\citealp{aerts2003a}) reports \textit{K}-values for $\beta\,$Cep stars which
are typically in the range of \numrange{0.01}{1}. Similar values are found in
our 2D simulation at frequencies of standing p-modes which is most easily seen
as sharp dips at frequencies above the \bvf. Such a simplified comparison of
values in the interior of our model to the observations at actual stellar
surfaces does not allow for a quantitative comparison or stronger conclusions.
However, it shows that the wave spectra that self-consistently arise in our
hydrodynamical simulation are compatible with observations of oscillations in
real stars of similar mass and evolutionary stage (i.e., SPB stars;
\citealp{decat2002b}).

\begin{figure}
  \centering
  \includegraphics[width=\columnwidth]{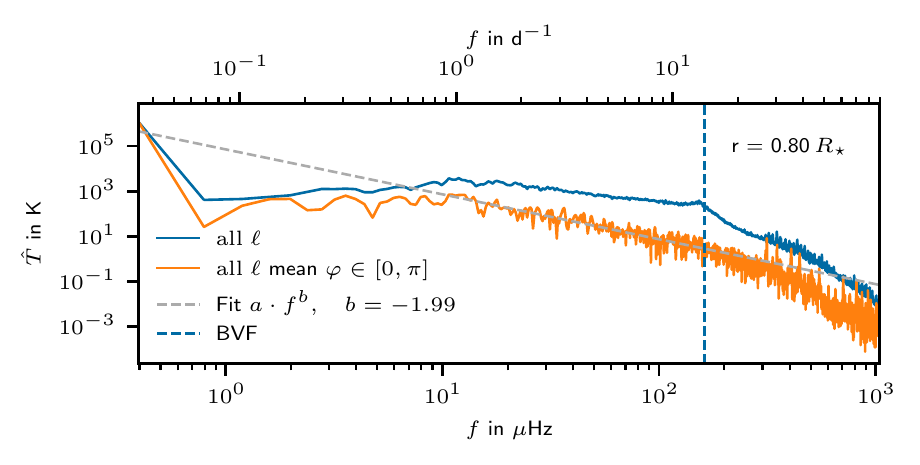}
  \caption{Spectra for temperature at $r=\SI{0.80}{R_\star}$. The orange
  line shows the spectrum of the temperature fluctuations which
  have been averaged over half of the circumference of our 2D
  model. The blue line corresponds to the spectra averaged over
  \num{100} radial rays after the FT as it is done for the
  velocity spectra. The dashed blue vertical line denotes the position of the
  \bvf. The gray line corresponds to a power-law fit for the 
  frequencies ranging from \SIrange{10}{200}{\micro\Hz} resulting in an
  exponent of $b \approx -2$.}
  \label{fig:line_mean}
\end{figure}

Besides surface velocities, the variation of the disk-averaged luminosity of
stars is another observable. It is supposed that contributions from small-scale
fluctuations (high $\ell$) are suppressed in this quantity as they cancel out
on average whereas large-scale contributions are pronounced \citep[e.g.,][Sect.
6]{aerts2010a}. As a simple estimate of this effect we compute the average
temperature over half of the circumference of our 2D model prior to the
temporal FT\@. The result is shown as an orange line in \cref{fig:line_mean}
(for comparison, we show as a blue line the spectra which have been averaged
similar to the velocity spectra, i.e., only after the temporal FT). At high
frequencies, pressure modes appear clearly as individual peaks. This is,
however, not the case in the low frequency regime.  We note that our spectrum
is like those presented in \citet{bowman2019a} (e.g., see their \figr3 for the
observables of the O star HD~46150 and the figures in their appendix
for additional OB stars). The actual situation for observations is more
complex than the diagnostic value of our simulations can be. For example, we
do not account for limb darkening while this phenomenon comes into
play to determine the net flux variation. The limb darkening has more
effect for flux observations but far less for the net velocity
variations in the line-of-sight. However, our simple experiment
illustrates that the net flux variations do not easily reveal any
distinct peaks corresponding to low-$\ell$ modes in the low-frequency
regime despite the expected cancellation of the numerous high-$\ell$
modes.

\subsubsection{Dispersion for pressure waves}\label{subsubsec:pdisp}

For the pressure waves, we perform a similar analysis as described in the
previous section except that we now check for the usual dispersion relation of
pressure waves
\begin{align}
  \frac{\omega}{c_{\text{sound}}} = \left| \vec k \right|.
  \label{eq:dispersion_pm}
\end{align}

\begin{figure}
  \centering
  \includegraphics[width=\columnwidth]{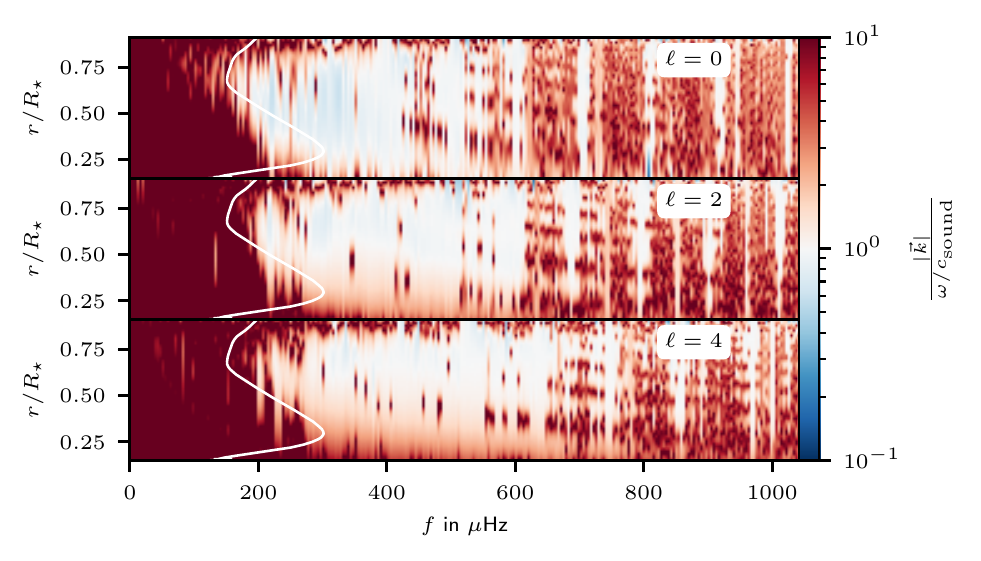}
  \caption{Same quantities as in \cref{fig:pcolor_dispersion_gm} but this time
  for the dispersion relation of sound waves (\cref{eq:dispersion_pm}).
  \label{fig:pcolor_dispersion_pm}}
\end{figure}

The result is shown in \cref{fig:pcolor_dispersion_pm} and the colors have the
same meaning as in \cref{fig:pcolor_dispersion_gm}.
As expected, the dispersion relation is not fulfilled for $f<N/2\pi$. At
frequencies corresponding to a standing mode we get a good agreement with
predictions from theory. Furthermore, \cref{fig:pcolor_dispersion_pm} indicates
that pressure modes are excited in the mixed-mode frequency regime, that is
between about \SI{200}{\mu\Hz} and \SI{300}{\mu\Hz}. This suggests that a
coupling of p- and g-modes is possible. The deviation between standing modes is
probably due to the same reasons as discussed in \cref{subsubsec:gdisp}. This
is further illustrated in \cref{fig:pdispersion_check} of the Appendix where we
show the amplitudes at two specific frequencies. When the frequency matches a
standing mode the wavelength can be detected easily whereas this is not
possible for frequencies in-between.  These results together with the fact that
the modes for a particular angular degree are equidistant in frequency in the
regime of high-order p-modes give us confidence that the observed modes are
indeed p-modes.

\subsection{Wave-amplification}\label{subsec:amplification}

From linear theory, one expects the amplification of IGW toward the surface
due to density stratification. As shown in \citet{ratnasingam2019a} for
spherical geometry, the prediction for IGW amplification is given by (in the
notation\footnote{We note that there is an error in the corresponding \eqr{23}
of \edel. The expression holds for the horizontal velocity component $v_h$
instead of the vertical velocity $v_r$ as stated.} of \edel)
\begin{align}
  v_{h} \propto
    \left(\frac{r_0}{r} \right)^{3/2} \sqrt{\frac{\rho_0}{\rho}}
    \left(\frac{N^2 - \omega^2}{N_0^2-\omega^2}\right)^{1/4}\exp(-\tau/2),
    \label{eq:amplification}
\end{align}
where $r_0$ is the starting point of the wave and $\rho_0$, $N_0$ the
corresponding density and \bvf. The damping factor $\tau$ is given by
\begin{align}
  \tau = \int_{r_0}^{r} \mathrm{d}r \frac{\kappa\,
  (\ell(\ell+1))^{3/2}N^3}{r^3\omega^4}\sqrt{1-\frac{\omega^2}{N^2}}.
\end{align}
According to \cref{eq:amplification}, waves are damped by the effects of
thermal dissipation and geometry, but amplified by decreasing density. This
ignores the damping effect of viscosity. For consistency,
\cref{eq:amplification} needs to be multiplied by a correction factor
$\left(r_0/r\right)^{-1/2}$ to account for our 2D polar geometry of an infinite
cylinder, which slightly increases the amplification
\citep[submitted]{ratnasingam2020a}. For the radial velocity,
\cref{eq:amplification} needs to be scaled according to \cref{eq:vratio}.
\begin{figure} 
  \centering
  \includegraphics[width=\columnwidth]{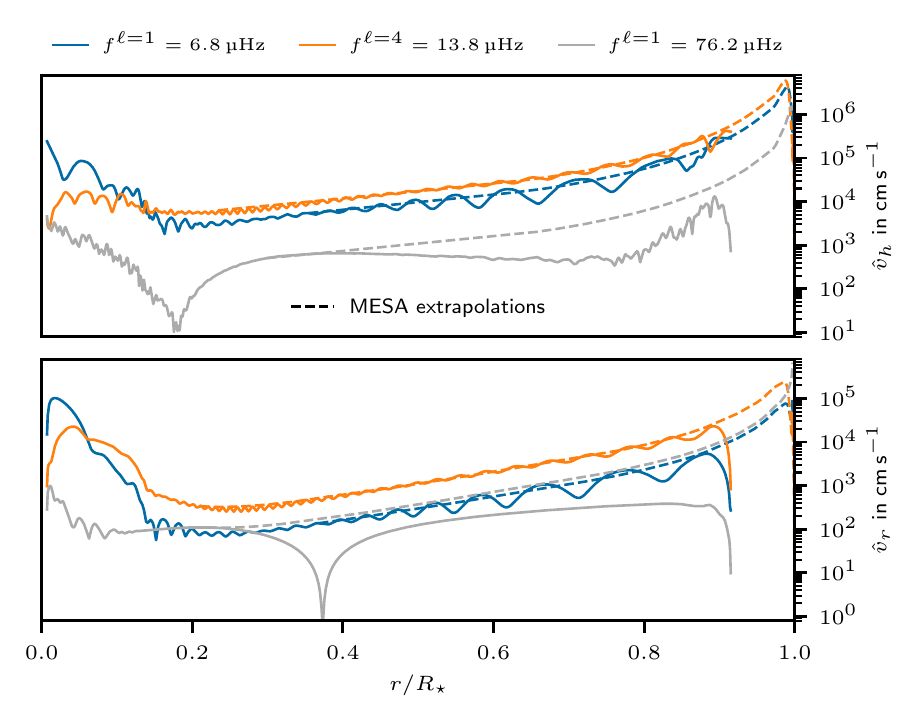}
  \caption{Velocity amplitude for different frequencies and angular degrees
  $\ell$ as function of radius. The amplitudes also include the
  contribution of the two adjacent frequency bins to account for the
  fact that peaks typically show a width of two to four bins in our
  simulation. The expected amplification toward the surface according
  to \cref{eq:amplification} is represented by dashed lines; the 1D
  MESA profile data serve as input variables whereas the starting
  points are taken as the simulated amplitudes at the first noticeable
  local maximum of the respective frequency and angular degree.}
  \label{fig:line_amplification}
\end{figure}

In \cref{fig:line_amplification} we compare the result of
\cref{eq:amplification} for the $\SI{3}{\msol}$ MESA model to the 2D
simulation, similar to \figr26 of \edel. For both radial and horizontal
velocity, the prediction from linear theory and the simulation are in good
agreement for short radial wavelengths and low frequencies ($f^{\ell = 1} =
\SI{6.8}{\mu\Hz}$, $f^{\ell = 4} = \SI{13.8}{\mu\Hz}$), and we assume that the
MESA extrapolations toward the surface provide a reasonable estimate of
surface velocities. At the highest frequency and longest wavelength
($f^{\ell=1} = \SI{76.2}{\mu\Hz}$) prediction and simulation clearly differ.
However, this is expected as the prediction assumes that wavelengths are short
compared to all relevant scale heights in the stratification which is clearly
not the case for the highest frequency shown in \cref{fig:line_amplification}.

Our extrapolation toward larger radii is a very simplified approach and we do
not account for the complex physics at the surface, such as the existence of a
subsurface convection zone.  Therefore, these numbers should only be seen as
an approximation to the order of magnitude of the velocity at the surface due
to amplified stellar oscillations. The drop in the simulation data at the very
top of the computational domain is an artifact of the numerical solid-wall
boundary condition whereas the drop in the prediction is an effect of stronger
radiative damping near the surface.

Also for the extrapolated velocities at the stellar surface, the order of
magnitude for the ratio of the horizontal velocity over the radial one shown in
\cref{fig:line_amplification} is compatible with what is found in time-series
spectroscopy of g-mode pulsators \citep{decat2002b}.

\subsection{Nonlinearity parameter}\label{subsec:nonlin_parameter}

If all waves were to remain in the linear regime, a treatment as in
\cref{subsec:amplification} would be a sufficient description of the physics
involved. Yet it is expected that the amplification toward the surface causes
nonlinearities to become relevant at least in certain ranges of frequency and
wavenumber. These nonlinearities lead to a redistribution of energy between
different wavenumbers and frequencies. Additionally, wave breaking can cause
transport of angular momentum from the core, where the waves are excited, to
the envelope.

In \citet{ratnasingam2019a} the effect of different input spectra on the
expected nonlinearity of IGWs is examined. The nonlinearity can be estimated by
\begin{align}
  \varepsilon = \frac{v_h}{\omega} k_h
  \label{eq:nonlin}
\end{align}
\citep{press1981a,barker2010a}. If $\varepsilon \gtrsim 1$, nonlinear effects
are dominant. However, as demonstrated in \citet[submitted]{ratnasingam2020a},
already rather small values of $\varepsilon$ ($\approx \num{e-3}$) can cause
noticeable energy transfer between different frequencies and wavenumbers.
\begin{figure}
  \centering
  \includegraphics[width=\columnwidth]{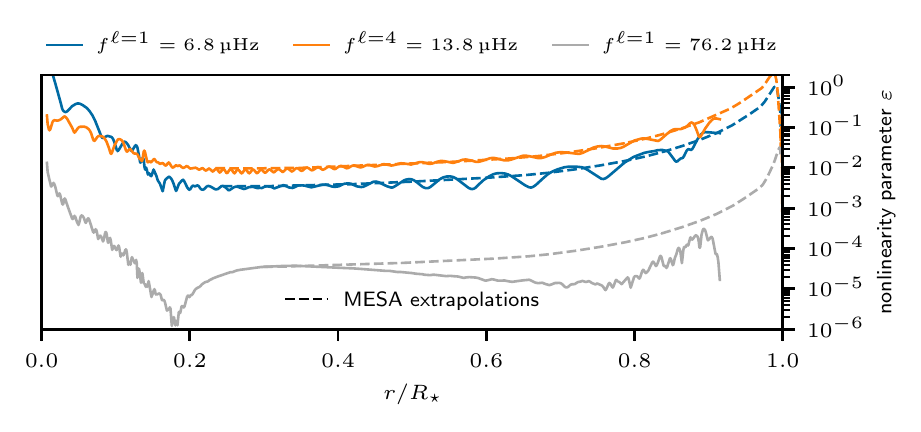}
  \caption{Nonlinearity parameter $\varepsilon$ according to
  \cref{eq:nonlin} for the same frequencies and angular degrees as in 
  \cref{fig:line_amplification}.}
  \label{fig:line_nonlinparam}
\end{figure}

In \cref{fig:line_nonlinparam} we show the result of \cref{eq:nonlin} for the
amplitudes of the horizontal velocity extracted from the simulation and the
extrapolation to the surface as illustrated in \cref{fig:line_amplification}.
The apparent increase in $\varepsilon$ with decreasing frequency is caused by
the stronger convective wave excitation at lower frequencies. At even lower
frequencies, which are not shown here, wave damping becomes dominant and
$\varepsilon$ decreases further. For the extrapolated amplitudes we find a
maximum value for the nonlinearity parameter of $\varepsilon = \num{2.2}$ for
$f^{\ell = 4} = \SI{13.8}{\micro\Hz}$.

\begin{figure}
  \centering
  \includegraphics[width=\columnwidth]{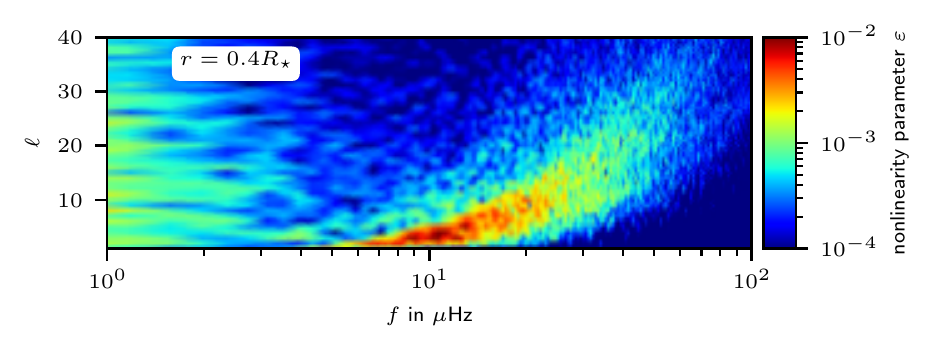}
  \caption{Nonlinearity parameter at $r=\SI{.4}{R_\star}$ for all available
  frequencies up to an angular degree of $\ell=40$.}
  \label{fig:pcolor_nonlinparam_4}
\end{figure}

From this we conclude that nonlinear effects can be expected at the surface in
the frequency regime around \SI{10}{\mu\Hz}.  This is further illustrated in
\cref{fig:pcolor_nonlinparam_4}, which shows the value for $\varepsilon$ at
$r=\SI{0.4}{R_\star}$ as a function of frequency $f$ and angular degree $\ell$.
A narrow range around \SI{10}{\mu\Hz} is apparent where the nonlinearity is
highest. We note that this looks very similar to \textit{Spectrum K} and
\textit{Spectrum LD} in \figr5 of \citet{ratnasingam2019a} for convective
velocity boosted by a factor of three with respect to the stellar models.  The
strongest nonlinearity in \cref{fig:pcolor_nonlinparam_4} is seen in the
frequency range in which gravity modes have been detected in a sample of about
30 SPB stars in the {\it Kepler\/} data. For all these pulsators, nonlinear
behavior has been deduced from the {\it Kepler\/} light curves and a
low-frequency IGW power excess has been detected after the removal of the
coherent g-modes \citep{pedersen2020a}.

We emphasize that the predictive power for the wave amplification and
nonlinearity from this \slh simulation is limited.  The boosted energy
generation in the core and the accompanying increased velocities also impact
the amplitudes of the excited waves.  Furthermore, convection is known to be
faster in 2D simulations and we apply a simple extrapolation from the upper
boundary of our 2D model toward the surface. Nevertheless, we believe that
the results presented here at least give an idea of what to expect at the
stellar surface where we find indication for the existence of nonlinear
effects.

\subsection{Time stepping \& efficiency}\label{subsec:timestepping}

After the detailed evaluation of the properties of the excited waves in the
previous sections, the paper is concluded with a discussion of the time step
sizes and the efficiency of the \slh code.

\begin{figure}
  \centering
  \includegraphics[width=\columnwidth]{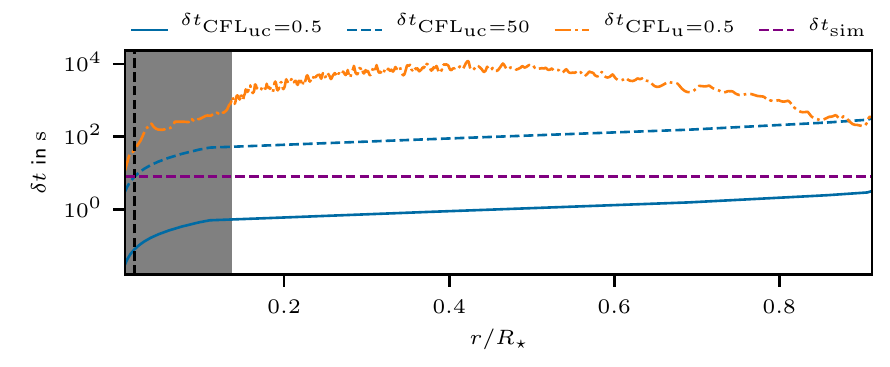}
  \caption{Time step size as a function of radius for different time step
  criteria. Because of the polar geometry of the grid, the innermost
  cells become narrower and thus require shorter time steps.  The
  purple dashed horizontal line denotes the time step size of $\delta t
  = \SI{8}{\second}$ that is used in the 2D simulation. The dashed
  vertical line marks the radius below which this step size is larger
  than a time step size according to \cref{eq:cfluc} with a CFL number
  of \num{50}. The gray shaded area indicates the convective region.}
  \label{fig:timesteps}
\end{figure}       
In \cref{fig:timesteps}, the radial profiles of the time step size
$\dtx{\cflx{uc}=0.5}(r)$ for explicit time stepping (see \cref{eq:cfluc}) and
the maximum implicit time step size $\dtx{\cflx{u}=0.5}(r)$ (see
\cref{eq:cflu}) are shown for values in the middle of the time span of the 2D
simulation. The time step sizes decrease toward the core due to the shrinking
azimuthal width of the cells in polar geometry. From the radial profiles, the
actual global time step is then given by the minimum over the whole domain. We
find $\min[\dtx{\cflx{uc}=0.5}(r)]=\SI{0.03}{\second}$ and
$\min[\dtx{\cflx{u}=0.5}(r)]=\SI{21.4}{\second}$. Hence, for implicit time
stepping, the possible time step size is roughly \num{700} times larger than
the corresponding explicit time step. This might be counter intuitive as the
Mach numbers at the surface are of order unity and one would therefore expect
implicit time steps to be closer to the explicit step sizes. In that sense,
implicit time stepping helps to overcome the general problem of very small step
sizes in polar and spherical geometries for large ratios of the outer to the
inner radius of the computational domain.

In \citet{miczek2013a}, the propagation of a simple 1D sound wave in \slh
is compared to predictions from linear wave theory. The simulations are
repeated while successively increasing the implicit time step size. In
this test setup, sound waves are resolved without noticeable
modification if $\delta t \leq \dtx{\cflx{uc}=5}$, whereas they get
considerably damped but still propagate at a speed close to the
theoretical prediction for $\delta t \approx \dtx{\cflx{uc}=50}$.  For
$\delta t \geq \dtx{\cflx{uc}=500}$ sound waves cannot be followed at
all anymore. 

For the 2D simulation presented here, these findings indicate that choosing a
time step size of $\dtx{\cflx{u}}=0.5$ which corresponds to $\delta t \approx
\dtx{\cflx{uc}=\num{350}}$ at the lower boundary might lead to a strong damping
of the sound waves. Although the appropriate time step size clearly depends on
the details of the setup and the wavelengths of the considered waves, we follow
\citet{miczek2013a} in choosing the time step size in this initial study. As
will be shown at the end of this section, this might give a value for the time
step size that is too conservative.

For the particular simulation presented here, the time step size is chosen to
be $\dtx{sim}=\SI{8}{\second}$ (dashed purple horizontal line in
\cref{fig:timesteps}). For radii smaller than the vertical line, sound waves
are expected to suffer from damping as for these cells the time step is larger
than $\dtx{\cflx{uc}=50}$. This, however, is only a tiny fraction of the whole
domain and we do not expect it to have a significant impact on the results. Our
choice is a compromise between efficiency and accuracy as will be demonstrated
in the following.

To quantify the gain in efficiency when using implicit time stepping, we
compare the wall clock time needed to cover a time span of \SI{80}{\minute}
(\num{2} convective turnover times) when using either implicit or explicit time
stepping. For this test, both simulations are restarted at $\SI{500}{\hour}$
and all output routines of \slh are disabled in order to minimize possible
external effects like a slow file system. We find that the explicit run
requires a wall clock time of \SI{37.72}{\minute} to perform \num{15900} steps
on \num{360} Intel Skylake cores. The implicit run finishes after
\SI{6.3}{\minute} while performing \num{600} time steps using the same number
of cores. Accordingly, the implicit run is roughly a factor of six faster than
the explicit run even though velocities at the top of the computational domain
are clearly not in the low-Mach regime, in which implicit time stepping is
usually advantageous.

We further compare the efficiency of the \slh code (implicit and explicit) to
the pseudo-spectral anelastic SPIN code (\edel).  For the comparison, the test
runs presented in this section are used for the \slh timings. For the same
physical problem, a short test simulation on 300 Ivy Bridge cores for a grid of
$1500(r)\times 128(\vartheta) \times 256(\varphi)$ serves as input for the SPIN
results. 

In \cref{tab:compare}, the timings for the two codes to perform one time step
per cell and core are listed. From this measure, the explicit \slh mode is
roughly a factor of five slower than the SPIN code, potentially reflecting the
fact that \slh has not undergone any substantial optimization effort. The
implicit \slh mode is \num{220} times slower than SPIN with a time step size
that is only eight times larger. These numbers illustrates that, while our
compressible simulations capture the full set of physics described by the full
Euler equations, they come with considerably higher computational costs.
Because the SPIN and \slh simulations differ in resolution,
computational domain, and the energy boosting, a quantitative comparison
considering the wall-clock time needed to evolve a cell by one unit of stellar
time is not possible here.

We note that the settings of the \slh implicit mode may not be
optimal yet. As stated by \citet{miczek2013a}, fine tuning the linear solver
for the specific physical problem to be solved and applying a multigrid solver
may lead to improved performance. This is subject of ongoing work and the
necessary algorithms are readily available in \slh.

\begin{table}[htpb]
  \centering
  \caption{Timings for the \slh code (implicit and explicit)
  and the pseudo-spectral anelastic code SPIN to perform one time step
  per cell and core.}
  \label{tab:compare}
  \begin{tabular}{lc}
    \toprule
    {\shortstack[l]{Code \\ \phantom{a}}}  &
    {\shortstack[c]{time$/$cell$/$step \\ $\left[\si{w-\micro\second}\right]$}} \\
    \midrule 
    \slh (impl)  & $\num{3.3e2}$ \\
    \slh (expl)  & \num{7.3}     \\
    SPIN         & \num{1.5}     \\ 
    \bottomrule
  \end{tabular}
\end{table}

To put the computational costs of the \slh runs into context of 3D simulations,
we have performed first preliminary low-resolution 3D simulations of the same
initial stratification as used for the 2D simulation in this paper on the
JUWELS supercomputer in J\"ulich, Germany. The domain is discretized
in $280(r)\times 90(\vartheta)\times 180(\varphi)$ cells while performing time
steps of $\delta t = \SI{8}{\second}$ with the \mbox{ESDIRK23} implicit time
stepping scheme. Based on these runs, we estimate that \SI{8e5}{\coreh} will be
needed to cover \SI{700}{\hour} of simulation time. By scaling this number with
the number of grid cells, a simulation with $480 \times 180 \times 360$ cells,
which corresponds to half of the resolution of the 2D simulation presented
here, needs \SI{5.5e6}{\coreh}. The reference simulation applies the same time
step size as the finer resolved 2D simulation; therefore, a possible change in
the step size is not included in the scaling. However, depending on the desired
accuracy, a much larger time step can be chosen and the computational costs
decrease correspondingly.

This demand of computational resources is a realistic scenario for applications
at HPC facilities. A 3D simulation at the same resolution as the 2D simulation
requires \SI{44e6}{\coreh} which is at the upper limit of common computing time
proposals. However, these numbers show that \slh is efficient enough to perform
simulations of stellar oscillations also in 3D for sufficiently long time
spans.

As demonstrated, the implicit time stepping improves the efficiency of the \slh
code. At the same time it is important to confirm that the accuracy at which
sound waves are evolved is sufficient. To this end, we restart the simulation
from the 2D implicit run at \SI{500}{\hour} and evolve it using an explicit,
three-stage Runge-Kutta (RK3) scheme \citep{shu1988a} which is third-order
accurate in time. The time step is set to $\delta t_{\cflx{uc}=0.5}$, our
standard choice for explicit time stepping.  The simulation is evolved for
\SI{100}{\hour} of physical time which corresponds to roughly \num{80} sound
crossing times $\tcross$. We define the sound crossing time to be the time a
sound wave needs to travel from the innermost point of the computational domain
to the upper boundary,
\begin{align}
  \tcross = \int_{r_0}^{r_1} \frac{1}{c_{\mathrm{sound}}(r)} \,\mathrm{d}r = \SI{1.26}{\hour}\,.
\end{align}
We expect the time span of \SI{80}{\tcross} to be sufficient to reveal possible
deviations in the spectra.

\begin{figure}
  \centering
  \includegraphics[width=\columnwidth]{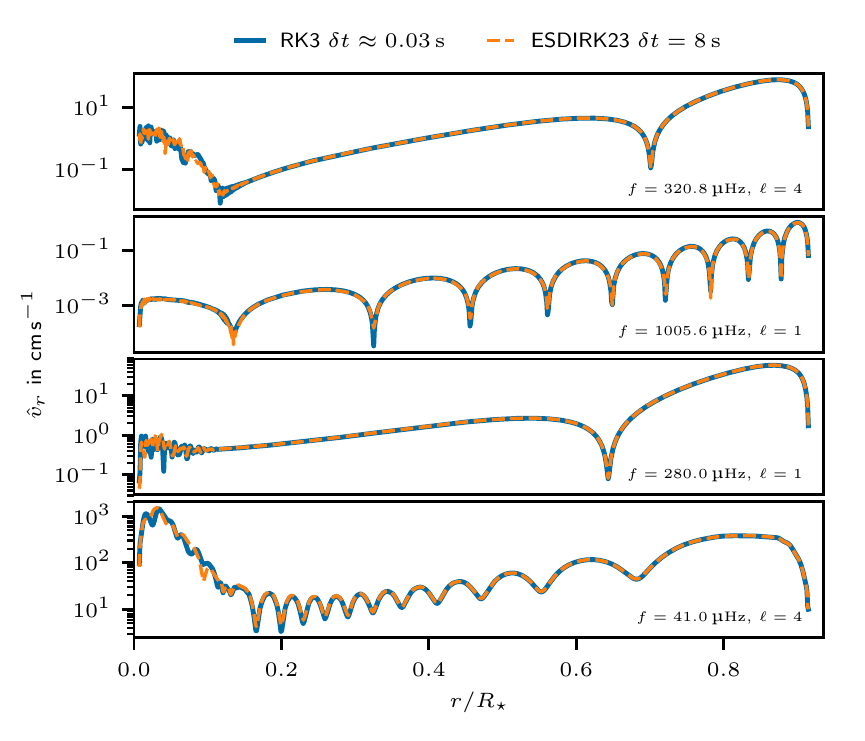}
  \caption{Amplitudes for the radial velocity spectra at four different
  frequencies for explicit (blue) and implicit (dashed orange) time stepping. The
  first two frequencies correspond to p-modes whereas the lower two frequencies
  match g-modes. For radii above the convection zone, the amplitudes completely
  overlap each other.}
  \label{fig:impl_expl}
\end{figure}

In \cref{fig:impl_expl}, the spectra of the implicit and explicit run are
compared at different frequencies corresponding to g- and p-modes. All spectra
are based on the same time frame. 

Almost no differences between the amplitudes from implicit and explicit time
stepping are visible. This result confirms that the propagation of sound waves
is treated correctly also in the implicit run. Furthermore, our implicit time
step size of \SI{8}{\second} appears to be a rather conservative choice and
larger steps sizes might be possible. This could reduce the computational costs
considerably and will be investigated in more detail in future simulations.

\section{Conclusion}\label{sec:conclusion}

The main goal of this paper has been to verify the capabilities of the
time-implicit, compressible \slh code to correctly treat internal gravity
waves (IGWs) and p-modes in stars with a convective core. To this end, two
test cases have been considered.

The first test is a simple 2D setup where an IGW packet according to the
Boussinesq approximation is evolved in time.

We first simulated the propagation of an IGW packet in a weakly-stratified 2D
atmosphere. The initial IGW was set up with different expected group
velocities. Velocity distributions were extracted from simulations performed
using the low-Mach \ausmpup solver and the classical \roe at a few points in
time. The ability of the two solvers to follow IGWs was assessed by comparing
the numerical solutions with approximate analytic solutions. The \roe revealed
strong damping and different broadening of the initial wave packet and the
packet's group velocity was incorrect. These effects were most pronounced at
Mach numbers $\gma=\num{e-3}$. The \ausmpup scheme showed no significant
damping and propagated the packet at a speed close to the prediction. This
indicates that a specialized solver is necessary when treating IGWs at low Mach
numbers.

Our second test case involving low-Mach-number flows was core hydrogen burning
in a \SI{3}{\msol} ZAMS star. We used the same initial 1D model as
\citet{edelmann2019a} (\edel) to make the simulations comparable. A 2D
simulation of this setup was performed using the \ausmpup solver. The
properties of IGW and p-modes were studied following similar methods as in
\edel. It was shown that the spectra extracted from the 2D \slh simulations
reflect the basic properties of internal waves.  A broad spectrum of IGWs is
observed for the integrated spectrum whereas individual standing modes can be
identified in spectra for single angular degrees $\ell$. Modes below the
Brunt--V\"ais\"al\"a frequency (\bvf) have an increasing radial order
with decreasing frequency, a fundamental property of IGWs\@. Also the
dispersion relation extracted from the simulation and the ratios of vertical to
horizontal velocities match the theoretical prediction. Furthermore, we find
the velocity ratios to be compatible with observational diagnostics from
time-series space photometry and high-resolution spectroscopy of slowly
pulsating B stars \citep{decat2002b}.

For standing modes above the \bvf, the radial order increases with increasing
frequency and the dispersion relation matches the one of p-modes. This kind of
waves cannot be seen in the anelastic approximation as sound waves are removed
from the underlying equations. 

Recently, \citet{lecoanet2019a} argued that if the observed variability of
stellar surfaces was due to the excitation of IGW from core convection, one
would expect to observe  distinct peaks in the spectrum which correspond to low
$\ell$ values (see their \figr2). We do not see such features in our simulation
(see, e.g., \cref{fig:line_mean}). Our explanation for the broad and
near-flat profile of IGWs is the same as in \edel: we considered the
entire ensemble of waves with large range in radial order and
$\ell$-values, resulting in small spacings between the resonant
frequencies. This ``hides'' frequency peaks due to individual
low-$\ell$ modes. We clearly see the individual resonant mode frequencies
showing up in the simulations when we limit to particular $\ell$-values, as
illustrated in \cref{fig:pcolor_lall012,fig:pcolor_lall012_h}. Moreover,
stellar rotation, which is not included in the simulations presented here,
would cause spectral line splitting and a further increase in the number of
lines in the spectrum.

The amplification of the waves toward the surface agrees with the expectation
given in \citet{ratnasingam2019a}. In \slh, the treatment of IGWs at very low
frequencies is mainly limited by radial resolution. In contrast, anelastic
simulations are limited by radiative damping and viscous effects already at
larger frequencies. Irrespective of the numerical setup, this work and that of
\edel demonstrate the importance of the excitation and propagation of IGWs as a
diagnostic tool for the interior physics of stars burning hydrogen in a
convective core.

The simulation of the \SI{3}{\msol} model presented here is intended as a proof
of concept and aids in the comparison of the simulations of \citet{rogers2013a}
and \citet{edelmann2019a}. The chosen 2D geometry reduces computational costs
and allows for parameter exploration. A validation of 2D results based on
selected 3D models is planned for future work. From the cost of
preliminary low-resolution 3D simulations we estimated a need of
\SI{44e6}{core\text{-}h} to simulate a grid with a size of $960 (r) \times 360
(\vartheta) \times 720 (\varphi)$ for \SI{700}{\hour} physical time. Using only
half of the number of cells in each dimension reduces the estimated cost to
\SI{5.5e6}{core\text{-}h}. These estimates are based on a reference run with
the same time step size as the higher resolved 2D simulation presented here.
Thus, a change in the time step size is not considered in the
scaling. Our tests indicate that the implicit time step size could
be increased while still resolving sound waves accurately enough. This could
considerably reduce the computational costs. Finally, \slh has already proven
an excellent scaling up to a large number of cores (e.g.,
\citealp{edelmann2016b}) such that this kind of 3D simulations are feasible on
modern HPC facilities.

After having tested the methods on the \SI{3}{\msol} model, we will extend our
study to higher stellar masses and later evolutionary stages for which there
are more observational data. Furthermore, we aim to use the velocity and
temporal information from our hydrodynamics data to extract synthetic
observables by averaging appropriately over the different wavenumber
components. Studying the dependence of wave amplitudes on different
luminosity boosting will help us to estimate the amplitudes by
extrapolating toward stellar values.

\begin{acknowledgements}
LH, FKR, and RA acknowledge support by the Klaus Tschira Foundation. PVFE was
supported by STFC grant ST/L005549/1 and by the US Department of Energy through
the Los Alamos National Laboratory. Los Alamos National Laboratory is operated
by Triad National Security, LLC, for the National Nuclear Security
Administration of U.S. Department of Energy (Contract No. 89233218CNA000001).
This work has been assigned a document release number LA-UR-20-24176. The
research leading to these results has received funding from the European
Research Council (ERC) under the European Union's Horizon 2020 research and
innovation programme (grant agreement No. 670519: MAMSIE). The authors
gratefully acknowledge the Gauss Centre for Supercomputing e.V.
(www.gauss-centre.eu) for funding this project by providing computing time
through the John von Neumann Institute for Computing (NIC) on the GCS
Supercomputer JUWELS at Jülich Supercomputing Centre (JSC). LH thanks Johann
Higl for insightful comments. We thank Daniel Lecoanet for useful remarks on
the paper and the constructive discussion. We also thank the anonymous referee
for very helpful and constructive comments that significantly improved this
paper.
\end{acknowledgements}

\bibliographystyle{aa}

\begin{appendix}

\section{Linear theory in Boussinesq approximation}\label{appendix:busigw}

In order to get analytical expressions describing the behavior of IGWs, the fully
compressible Euler equations need to be linearized. This can be done in the
Boussinesq approximation. It is based on the assumption that pressure and
density vary only little in the volume considered. Furthermore, one imposes a
time-independent hydrostatic background state and only follows the temporal
evolution of deviations from the background state.  Additionally, a
divergence-free velocity field is assumed. This approach removes the physics of
sound waves. For a Boussinesq gas it is convenient
to introduce potential temperature $\vartheta$ as
\begin{align}
  \vartheta = T\left(\frac{p}{p_0} \right)^{-(\gamma-1)/\gamma},
  \label{eq:ptemp}
\end{align}
where $p_0$ is the pressure at a specific reference height in the atmosphere
with $\vartheta_0,\ \rho_0$ (see \citealp{sutherland2010} for a detailed
introduction). The two-dimensional equations of motions can then be written as
\begin{align}
  \frac{\DD\tilde{\vartheta}}{\DD t} + v\frac{\dd\thse}{\dd y}
     &=0, \label{eq:b1}\\
  \frac{\DD u}{\DD t} + \frac{1}{\rho_0}\frac{\partial \tilde{p}}{\partial x}
     &=0, \label{eq:b2}\\
  \frac{\DD v}{\DD t} + \frac{1}{\rho_0}\frac{\partial \tilde{p}}{\partial y}&=
     -\frac{g}{\vartheta_0}\tilde{\vartheta}, \label{eq:b3}\\
  \frac{\partial u}{\partial x} + \frac{\partial v}{\partial y} &= 0,
     \label{eq:b4}
\end{align}
where quantities with a tilde denote fluctuations from the hydrostatic
background state, for instance $\tilde p = p - p_{\mathrm{hse}}$. The letters $u,\ v$
refer to the horizontal and vertical components of the velocity $\vec v = (v,u)^{T}$.
Further, the notation above makes use of the material derivative $\DD q / \DD t
= \partial q/\partial t + \vec v \nabla q$.

A solution to this set of equations can be found using the ansatz of a 2D plane
wave
\begin{align}
  \begin{pmatrix}
    \tilde\vartheta \\ u \\ v \\ \tilde p
  \end{pmatrix} =
  \begin{pmatrix}
    A_\vartheta \\ A_u\\ A_v \\ A_p
  \end{pmatrix}
  \, \exp{\left[i(\vec k\cdot \vec x) - i\omega t\right]},
  \label{eq:2Dwave}
\end{align}
which introduces the wave vector $\vec k = (k_x, k_y)^{T}$, the angular velocity
$\omega$, and the complex amplitudes $A_i$. The absolute values of these
amplitudes are assumed to be small, such that terms with products of two or more
amplitudes can be neglected. This essentially removes the advection term in the
material derivative. Inserting the ansatz into \cref{eq:b1,eq:b2,eq:b3,eq:b4}
results in a homogeneous system of linear equations of the form
\begin{align}
  \begin{pmatrix}
    -i\omega & 0 & \frac{\dd\thse}{\dd y} & 0                  \\
    0        & -i\omega & 0 & \frac{ik_x}{\rho_0}              \\
    \frac{g}{\vartheta_0} & 0 & -i\omega & \frac{ik_y}{\rho_0} \\
    0 & ik_x & ik_y & 0
  \end{pmatrix}\cdot
  \begin{pmatrix}
    A_\vartheta \vphantom{\frac{\dd\thse}{\dd y}} \\
    A_u\vphantom{\frac{ik_x}{\rho_0}}             \\
    A_v\vphantom{\frac{ik_x}{\rho_0}}             \\
    A_p\vphantom{ik_x}
  \end{pmatrix}
  =M \cdot \vec A = \vec 0.
  \label{eq:lsys}
\end{align}
Nontrivial solutions exist only if $\det(M) = 0$ which leads to the dispersion
relation for Boussinesq IGWs
\begin{align}
  \omega^2 &= -\frac{\dd\thse}{\dd y}\,\frac{g}{\vartheta_0}\,
    \frac{k_x^2}{|\vec k|^2}
  = -\frac{\dd\thse}{\dd y}\,\frac{g}{\vartheta_0}\,
    \cos^{2}(\theta),
  \label{eq:bdisp1}
\end{align}
where we have used $\vec k\cdot \vec e_x = k_x = |\vec k|\cos\theta$ with $\vec
e_x$ being the unit vector in horizontal direction and $\theta$ the angle
between the wave vector $\vec k$ and the horizontal direction. The dispersion
relation is usually written in the form
\begin{align}
  \omega = N_0 \cos(\theta),
  \label{eq:bdisp}
\end{align}
where
\begin{align}
  N_0 &= \sqrt{-\frac{\dd\thse}{\dd y}\, \frac{g}{\thse}}
  \approx\sqrt{-\frac{\dd\thse}{\dd y}\, \frac{g}{\vartheta_0}}
  \label{eq:bvf}
\end{align}
is the \bvf in the Boussinesq approximation (see, e.g., \citealp{sutherland2010},
Sect.~3.2). This result shows that the angular frequency does not depend on the
absolute value of the wave vector $\vec k$ and that IGWs do
not propagate isotropically. The maximum frequency is $\omega = N_0$ for
$\theta=\SI{0}{\degree}$ and no purely vertical waves exist as
$\omega=\SI{0}{}$ for $\theta=\SI{90}{\degree}$.

For the specific solution of \cref{eq:lsys}, we set the amplitude of the
vertical velocity as free parameter and express the other amplitudes
accordingly:
\begin{align}
  A_\vartheta &= -\frac{i}{\omega}\frac{\dd\thse}{\dd y}\, A_v, \label{eq:At} \\
  A_u &= -\frac{k_y}{k_x}\, A_v, \label{eq:Au}                                \\
  A_p &= -\rho_0\omega \frac{k_y}{k_x^2}\, A_v. \label{eq:Ap}
\end{align}
As \cref{eq:lsys} is a linear system, any superposition of solutions remains a
solution to the system. The group velocity of such a wave packet is then given by
\begin{align}
  \vec c_g = \nabla_{\vec k} \omega = \frac{N_0}{k_x}\cos\theta\sin\theta
  \begin{pmatrix}
    \sin\theta \\ -\cos\theta
  \end{pmatrix},
  \label{eq:gvel}
\end{align}
if one uses $\sin\theta = k_y / |\vec k|$.

\subsection{Evolution of a wave packet} \label{subsec:packetevo}

The following section is based on the methods described in
\citet[Sect.~1.15]{sutherland2010} which we generalize to 2D and apply to the
specific setup presented in \cref{sec:boussinesq}.

In linear theory, a quasi-monochromatic wave packet $\eta(\vec x, t)$ is
usually described as
\begin{align}
  \eta(\vec x, t) &= \Aa(\vec x, t)\exp\left[ \ii \left(\vec k_0\cdot\vec x - \omega\left( \vec k_0 \right) t\right)\right],
  \label{eq:etax}
\end{align}
where the amplitude modulation function $\Aa(\vec x, t)$ changes much slower in
space and time than the exponential function in \cref{eq:etax}. The wave
packet $\eta(\vec x, t)$ can be represented via a FT as
\begin{align}
  \eta(\vec x, t) &= \int_{-\infty}^{\infty}\hat{\eta}(\vec k)
    \exp\left[\ii\left(\vec k\cdot \vec x - \omega\left(\vec k\right)t \right)\right]  \, \dd \vec k
  \label{eq:etak}
\end{align}
and, by applying the inverse FT, its spectral representation reads
\begin{align}
  \hat\eta(\vec k) &= \frac{1}{(2\pi)^2}\int_{-\infty}^{\infty} \eta(\vec x, 0) \exp\left[-\ii\vec k\cdot \vec x\right]\, \dd\vec x.
  \label{eq:etaxft}
\end{align}
From \cref{eq:etax} and \cref{eq:etaxft} it further follows that 
\begin{align}
  \Aa(\vec x, t) &= \eta(\vec x,t) \exp\left[-\ii\left(\vec k_0\cdot\vec x - \omega\left(\vec k_0\right) t\right)\right] \nonumber\\
  &= \int_{-\infty}^{\infty}\hat{\eta}(\vec k)\exp\left[\ii\Dq{\vec k}\cdot \vec x\right] 
  \exp\left[-\ii\Dq{\omega} t\right]  \, \dd \vec k,
  \label{eq:ampx}
\end{align}

where $\Dq{\vec k} = \vec k - \vec k_0$ and $\Dq \omega = \omega\left( \vec k
\right)- \omega\left(\vec k_0\right)$.  For $t=0$, \cref{eq:ampx} illustrates
that an initial amplitude modulation $\Aa(\vec x, 0)$  introduces waves with
$\vec k \neq \vec k_0$. Because the wave packet is quasi-monochromatic with a
typical wavenumber of $\vec k_0$, the amplitude $\hat\eta(\vec k)$ must
decrease quickly for $\vec k \neq \vec k_0$. However, in the presence of a
nontrivial dispersion relation $\omega = \omega\left( \vec k \right)$, any
superposition of waves with different wavenumbers will lead to a broadening of
the initial wave packet over time. This will be explicitly shown for the
setup presented in \cref{sec:boussinesq}.

Here, the initial vertical velocity modulation is a Gaussian as given by
\cref{eq:vamp}:
\begin{align}
  \eta(\vec x, 0) &= \underbrace{\Aa_0 \exp\left[-\frac{y^2}{2\sigma^2}\right]}_{\Aa(\vec x, 0)}
    \exp\left[\ii\vec k_0\cdot \vec x\right]
\end{align}
with 
\begin{align}
  \Aa_0 &= \fma\,\sqrt{\gamma\, \mathcal{R} T_0/\mu} ; \quad \sigma = \beta H_p/2.
\end{align}
Evaluating the FT in \cref{eq:etak} leads to
\begin{align}
  \hat\eta(\vec k) &= {\delta}(k_x-k_{x,0})\sqrt{2\pi}\sigma \Aa_0 \exp\left[-\frac{1}{2}\sigma^2 \Dq{k_y}^2\right],
  \label{eq:etakgauss}
\end{align}
where $\delta$ denotes the Dirac delta function. This specific form of
$\eta(\vec k)$ illustrates that a narrower Gaussian modulation in real space
leads to a broader distribution in wavenumber space and consequently to a
larger dispersion.

From \cref{eq:etakgauss,eq:ampx} the time evolution of $\Aa$ is determined. To
evaluate the FT in \cref{eq:ampx}, we expand the dispersion relation in a
Taylor series up to second order
\begin{align}
  \omega(\vec k)
  &=\omega(k_0) + \nabla_{\vec k}\omo \Dq{\vec k} 
   + \frac{1}{2}\left(\partial^2_{k_x,k_x}\omo\Dq{k_x}^2 + \partial^2_{k_y,k_y}\omo\Dq{k_y}^2\right) \nonumber\\
  &\phantom{=} + \partial^2_{k_x,k_y}\omo\Dq{k_x}\Dq{k_y} + \mathcal{O}\left( k^3 \right)
  \label{eq:omtaylor}
\end{align}
where we have introduced the abbreviations $\partial_q =
\frac{\partial}{\partial q},\, \partial_{p,q}^2 = \frac{\partial^2}{\partial p\partial q}$
for convenience. Inserting \cref{eq:omtaylor,eq:etakgauss} into \cref{eq:ampx} gives
\begin{align}
  &\Aa(\vec x, t) = \sqrt{2\pi}\sigma\Aa_0\,\int_{-\infty}^{\infty}\exp\left[\ii \Dq{k_y} y\right] \nonumber\\ 
  &\quad\cdot\exp\left[-\frac{1}{2}\sigma^2\Dq{k_y}^2
  -\ii\left(\nabla_{\vec k}\omo + \partial^2_{k_y,k_y}\omo\Dq{k_y^2}\right)t \right]\,\dd k_y.
\end{align}
This can be simplified to 
\begin{align}
  &\Aa(\vec X, t) = \sqrt{2\pi}\sigma\Aa_0 \int_{-\infty}^{\infty}
    \exp\left[-\frac{\tilde\sigma^2}{2}\Dq{k_y}^2\right]\exp\left[\ii\Dq{k_y} Y\right]\,\dd k_y,
  \label{eq:ampxs}
\end{align}
by transforming into a moving frame via 
\begin{align}
  \vec X = (X,Y)^{T} = \vec x - \nabla_{\vec k} \omot
\end{align}
and setting
\begin{align}
  \tilde\sigma = \sqrt{\sigma^2 + \ii\partial^2_{k_y,k_y}\omot}.
\end{align}
The integral in \cref{eq:ampxs} is again the FT of a Gaussian 
and leads to the final expression
\begin{align}
  \Aa(\vec X, t) &= \Aa_0\,\frac{\sigma}{\tilde\sigma}\exp\left[\frac{-Y^2}{2\tilde\sigma^2}\right]
  \label{eq:ampxf}
\end{align}
where the absolute value of $\Aa$ is given by
\begin{align}
  \left|\Aa(\vec X, t)\right| 
  &= \frac{\Aa_0 \sigma}{\left( \sigma^4+\left(\partial^2_{k_y,k_y}\omo t\right)^2 \right)^{1/4}} \cdot
  \exp\left[-\frac{Y^2}{2} \frac{1}{\sigma^2 + \left(\frac{\partial^2_{k_y,k_y}\omot}{\sigma}\right)^2}\right].
  \label{eq:aampxf}
\end{align}
\cref{eq:ampxf,eq:aampxf} describe the broadening and decrease in amplitude of
an initial Gaussian profile over time when considering the dispersion relation
up to second order in $\vec k$ in a system of coordinates moving with the wave
packet at its group velocity $\nabla_{\vec k} \omo$.  For the dispersion
relation of a Boussinesq IGW according to \cref{eq:bdisp}, the group velocity is
given by \cref{eq:gvel} and
\begin{align}
  \partial^2_{k_y,k_y}\omo = N_0 k_{0,x} \frac{2k_{0,y}^2-k_{0,x}^2}{\left( k_{0,x}^2+k_{0,y}^2 \right)^{5/2}}.
\end{align}

\section{Supplementary plots and tables}

\begin{figure}[h!]
  \centering
  \includegraphics[width=\columnwidth]{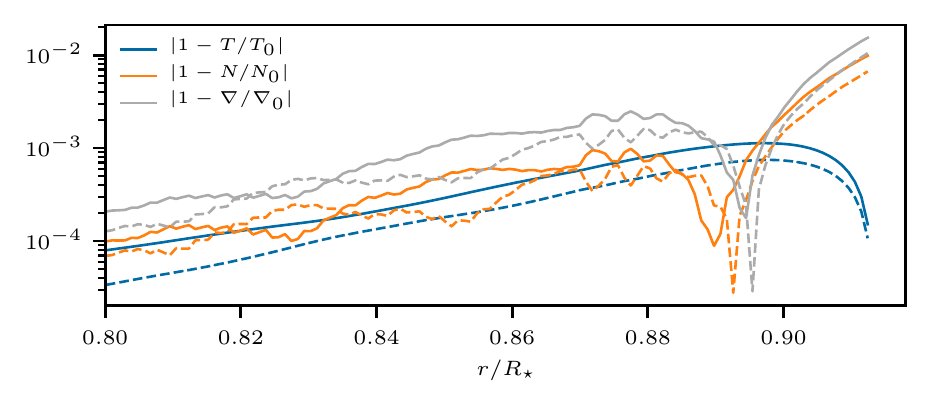}
  \caption{Relative deviation from the initial profile for the 1D
  simulations described in \cref{subsec:initmodel} at
  $t\sim\SI{1100}{\hour}$. Shown are results for the
  temperature $T$, the \bvf $N$, and the temperature gradient
  $\nabla = \partial\log T / \partial\log P$, respectively.
  Solid lines correspond to cylindrical geometry whereas dashed
  lines denote the results for spherical geometry. Only the
  surface of the computational domain is shown to emphasize the
  change at the outer radial boundary. The deviations are even
  smaller in the inner part which is not shown here.}
  \label{fig:compare1D}
\end{figure}

\begin{table}[htpb]
  \centering
  \caption{Maximum deviations for the quantities shown in
  \cref{fig:compare1D}. The subscripts denote cylindrical and spherical
  geometry, respectively.}
  \label{tab:compare1D}
  \begin{tabular}{lc}
    \toprule
    quantity & maximum                                                \\
    \midrule 
    $\left|1 - T/T_0\right|_\mathrm{cyl}$            & $\num{1.1e-3}$ \\ 
    $\left|1 - T/T_0\right|_\mathrm{sph}$            & $\num{7.5e-4}$ \\ 
    \midrule 
    $\left|1 - N/N_0\right|_\mathrm{cyl}$            & $\num{9.9e-2}$ \\ 
    $\left|1 - N/N_0\right|_\mathrm{sph}$            & $\num{6.7e-3}$ \\ 
    \midrule 
    $\left|1 - \nabla/\nabla_0\right|_\mathrm{cyl}$  & $\num{1.5e-2}$ \\ 
    $\left|1 - \nabla/\nabla_0\right|_\mathrm{sph}$  & $\num{1.1e-2}$ \\ 
    \bottomrule
  \end{tabular}
\end{table}

\begin{figure}[h!]
  \centering
  \includegraphics[width=\columnwidth]{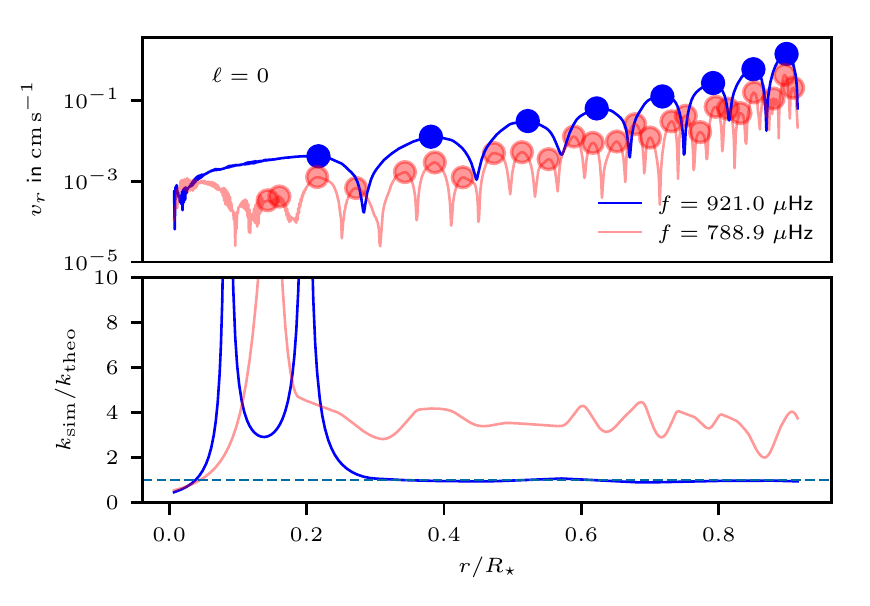}
  \caption{Check for the dispersion relation of sound waves.
  The \textbf{upper panel} shows the amplitudes of the radial velocity
  at two different frequencies for $\ell=0$. Circles mark radii where our
  routine detects peaks in the amplitude. The radial wavelengths are
  then estimated from the distance of neighboring peaks. For the blue
  line, the frequency matches a standing mode and it shows well defined,
  distinct peaks. The red line corresponds to a frequency which is
  between standing modes and shows small-scale incoherent oscillations
  with many peaks. In the \textbf{lower panel} the resulting wave
  numbers from the simulation are compared to the expectation for sound
  waves $k_\mathrm{theo} = 2\pi f/c_\text{sound}$.  We find the blue
  line in good agreement with theory for radii above the convection
  zone while the red line considerably differs.}
  \label{fig:pdispersion_check}
\end{figure}
\end{appendix}
\end{document}